\documentclass[%
reprint,
prb,
]{revtex4-2}
\usepackage{placeins}
\usepackage{graphicx}
\usepackage{dcolumn}
\usepackage{bm}
\usepackage{multirow}

\begin{document}
\title{
Effects of shear displacement on the conductance of monolayer/gapped bilayer/monolayer
graphene junctions: Implications for ac-dc conversion 
}

\author{Ryo Tamura}
\affiliation{
Faculty of Engineering, Shizuoka University, 3-5-1 Johoku, Hamamatsu 432-8561, Japan}

\date{\today}

\begin{abstract}
Analytical treatments of tunneling in bilayer graphene have typically relied on minimal models including only the vertical interlayer hopping $\gamma_1$ and have been restricted to the weak interlayer-bias regime ($2\varepsilon \ll \gamma_1$). Consequently, they cannot adequately describe lattice deformations or strong electric-field effects.
In this work, we present an analytical theory of evanescent states in electrically gapped bilayer graphene that overcomes both limitations.
Our approach explicitly incorporates the skew interlayer hoppings $\gamma_3$ and $\gamma_4$ and remains valid even when the interlayer bias $2\varepsilon$ is comparable to $\gamma_1$.
Focusing on low-energy electronic states near the charge neutrality point,
we analytically derive the complex longitudinal wave numbers, the gap width, and the sublattice pseudospin within the electric-field-induced gap.
We then systematically analyze the dependence of these quantities on the interlayer shear displacement $\vec{\delta}=(\delta_x,\delta_y)$, and 
 find that skew interlayer hoppings, in particular $\gamma_3$, play an essential role.
For transport along the zigzag ($x$) direction, the longitudinal wave vector becomes complex, whereas the transverse wave vector remains real.
For a monolayer/bilayer/monolayer junction with transport along the zigzag direction, we find that $\delta_y$ has a significantly stronger impact on the conductance than $\delta_x$.
Furthermore, we identify a shear-induced phase proportional to $\delta_y$ that appears universally in the analytical expressions for the gap width, the sublattice pseudospin, and the decay length.
These results establish a unified framework for shear- and bias-controlled evanescent tunneling in bilayer graphene and suggest broader relevance to nonequilibrium transport phenomena in layered materials.
\end{abstract}

\maketitle

\section{Introduction}
Quantum transport across heterojunctions is governed not only by propagating states but also by evanescent modes, whose complex wave vectors determine tunneling amplitudes and decay lengths inside barrier regions~\cite{Datta,T-Ando,CDL-1,CDL-2,pristine-bi-gap-1,pristine-bi-gap-2}.
A representative system in which evanescent modes play a central role is the monolayer/bilayer/monolayer graphene junction (MBMGJ), where 
 Bernal-stacked bilayer graphene forms the central region, while the outer monolayer regions act as the source and drain leads, as illustrated in Figs.~1(a) and 1(b).
Because a perpendicular electric field opens an energy gap exclusively in the central bilayer region, the junction realizes an atomically sharp tunneling barrier. The barrier height is determined by the gap width and can be controlled \emph{in situ} via the applied electric field~\cite{gra-Gap-1,gra-Gap-2,gra-Gap-3,gra-Gap-Ohta,gra-Gap-experiment-1,gra-Gap-experiment-2,gra-Gap-experiment-3,gra-Gap-experiment-4}. 
 Numerous theoretical studies have investigated the conductance of such MBMGJs within the Landauer formalism~\cite{up-Tamura,Tamura-2025,updown-Tamura,up-1,up-2,up-3,up-4,up-5,down-1,down-2,down-3,down-4,down-5,updown-1,updown-2,updown-3,updown-4,updown-5,updown-6}. 
However, most analytical approaches are restricted to the regime where the electrically induced gap is much smaller than the vertical interlayer hopping $\gamma_1$. When the gap becomes comparable to $\gamma_1$ and the Fermi level lies inside the gap, conventional low-energy reductions break down. Consequently, analytical treatments become scarce~\cite{pristine-bi-gap-1,pristine-bi-gap-2,pristine-bi-gap-3}.
While reduced two-band descriptions have proven useful in moir\'{e} graphene systems and related gauge-field analyses, these approaches typically focus on low-energy miniband physics~\cite{twist-1,twist-2,twist-3,twist-4,McCann,KoshinoMcCann2009,McCann2013-revier-reduced-model}. 
 In contrast, the present problem involves evanescent states inside an electrically induced gap comparable to $\gamma_1$, where the full four-band structure becomes essential.
Although Refs.~\cite{up-Tamura} and \cite{Tamura-2025} addressed this parameter regime, the microscopic origin of the shear-displacement dependence remained unclear.
One of the central aims of the present work is to clarify this issue.

Interlayer shear vibrations in bilayer and few-layer graphene have been observed experimentally by Raman spectroscopy and time-resolved optical measurements. 
These experiments have revealed low-frequency shear phonon modes and their dependence on the number of layers~\cite{Tan2012Shear,2013NanoLett,LCM}. 
Despite these experimental advances, analytical descriptions of the associated electronic effects remain relatively limited.
 By contrast, the theory of intralayer strain is well established, with lattice deformations described as effective gauge fields within continuum Dirac models~\cite{CastroNeto2009,Suzuura2002,Sasaki2006,Guinea2010,Vozmediano2010}. 
While this gauge-field framework has clarified the microscopic origin of 
electron--phonon coupling in monolayer graphene, interlayer shear 
displacements in bilayer systems have been explored far less extensively, 
especially when a large electric-field-induced energy gap is present~\cite{shear-theory-1,shear-theory-2}. First-principles calculations have examined the relationship between the gap width and shear displacement; however, the underlying physical origin has not yet been clarified~\cite{shear-theory-3}.

It is also worth noting that a variety of one-dimensional electronic states inside the bulk energy gap of graphene-based systems have been extensively studied. These include edge states~\cite{edge-0,edge-1,edge-2,edge-3,edge-4,edge-5} 
 and domain-wall states, where the domain walls arise from differences in perpendicular electric fields~\cite{domain-gate-0,domain-gate-1,domain-gate-2,domain-gate-3,domain-gate-4,domain-gate-5,domain-gate-6,domain-gate-7,domain-gate-8}, interlayer stacking~\cite{domain-gate-0,domain-stacking-2,domain-stacking-3,domain-stacking-4,domain-stacking-5,domain-stacking-6},  or sublattice site energies~\cite{domain-sublattice-1,domain-sublattice-2}. These states propagate along specific interfaces and are often protected by topology or symmetry.
In contrast, the present work focuses on evanescent bulk states inside the gap, which do not form real dispersion lines (RDLs) but instead govern tunneling and decay processes across gapped regions. The electronic structure of such evanescent states is naturally described in terms of complex dispersion lines (CDLs), in which the real and imaginary parts of the wave vector encode oscillatory and decaying behavior, respectively.
While CDLs have been extensively studied in
conventional semiconductor heterostructures~\cite{CDL-1,CDL-2},
studies of CDLs and evanescent modes in graphene-based
systems have mostly been limited to models that neglect
the skew interlayer hoppings 
$\gamma_3$ and $\gamma_4$~\cite{g-CDL-1,g-CDL-2,g-CDL-3,g-CDL-4}.
These skew hoppings play a more significant role than
the vertical hopping $\gamma_1$ in determining the
influence of interlayer shear displacement on the CDLs.

In this work, we investigate the complex dispersion lines (CDLs)
of Bernal-stacked bilayer graphene under a perpendicular electric field
using a tight-binding model that fully incorporates the interlayer
hopping parameters and interlayer shear displacement.
We focus on low-energy electronic states near the charge neutrality point
while allowing for a large interlayer bias comparable to the dominant
interlayer hopping $\gamma_1$.
We derive approximate analytical expressions for the CDLs that
accurately reproduce exact numerical results over a wide energy range
inside the gap and clarify their physical origin in terms of
shear-modified interlayer coupling.
These results are then applied to analyze electron transmission
across MBMGJs, where the shear-induced modification
of the decay length directly controls the conductance.
Finally, we discuss a possible experimental setup in which dynamically driven shear displacement, combined with an external probe field, may generate a dc current. This proposal provides a direct way to explore nonequilibrium tunneling governed by evanescent states.

\begin{figure}
\begin{center}
\includegraphics[width=\linewidth]{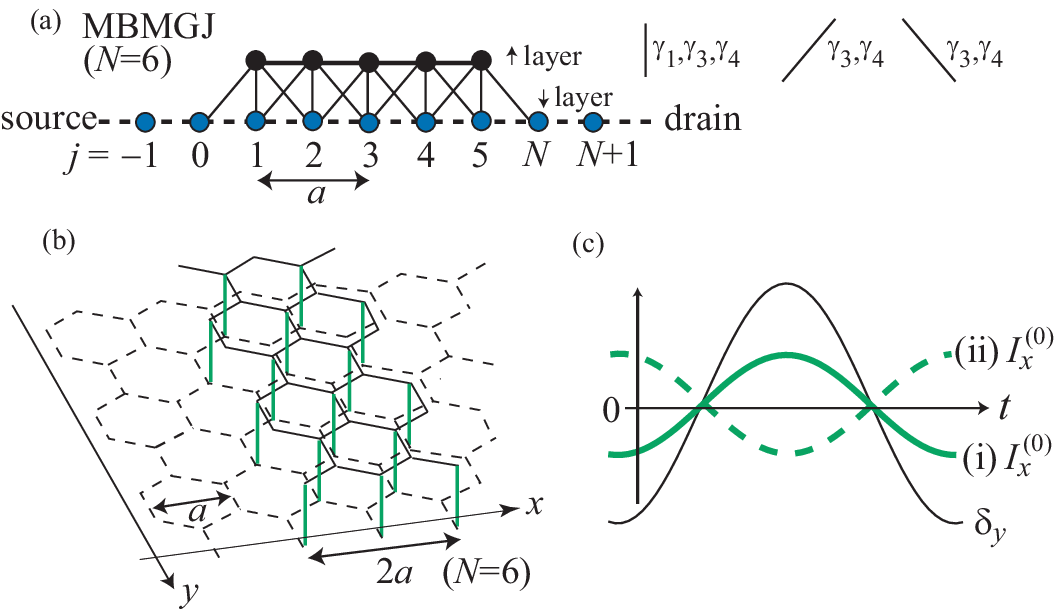}
\caption{
(a) 
Side view of a monolayer/bilayer/monolayer graphene junction (MBMGJ).
The $\uparrow$ layer of length $(N-2)a/2$ is Bernal-stacked on the $\downarrow$ layer, which is connected to the source and drain electrodes,
where $N$ is an integer and $a$ is the lattice constant.
The vertical interlayer bonds include $\gamma_1$, $\gamma_3$, and
$\gamma_4$, which overlap in this side view, whereas the inclined
interlayer bonds include $\gamma_3$ and $\gamma_4$.
The distinction among these hopping parameters can be seen more
clearly in Fig.~2(c).
(b) Schematic three-dimensional view.
(c) Conceptual illustration of a pump--probe scheme:
a $y$-polarized pump field, arriving first and independently controlled,
  induces a shear displacement $\delta_y$ along the $y$ direction,
while a delayed, independently controlled $x$-polarized probe field induces the probability current $I_x^{(0)}$ along the $x$ direction for electrons incident from the monolayer region.
Cases where $\delta_y$ and $I_x^{(0)}$ are (i) in phase and (ii) in antiphase are shown.
The horizontal axis represents time $t$.
See Sec.~VIII for details.}
\end{center}
\end{figure}
\begin{figure*}
\begin{center}
\includegraphics[width=\linewidth]{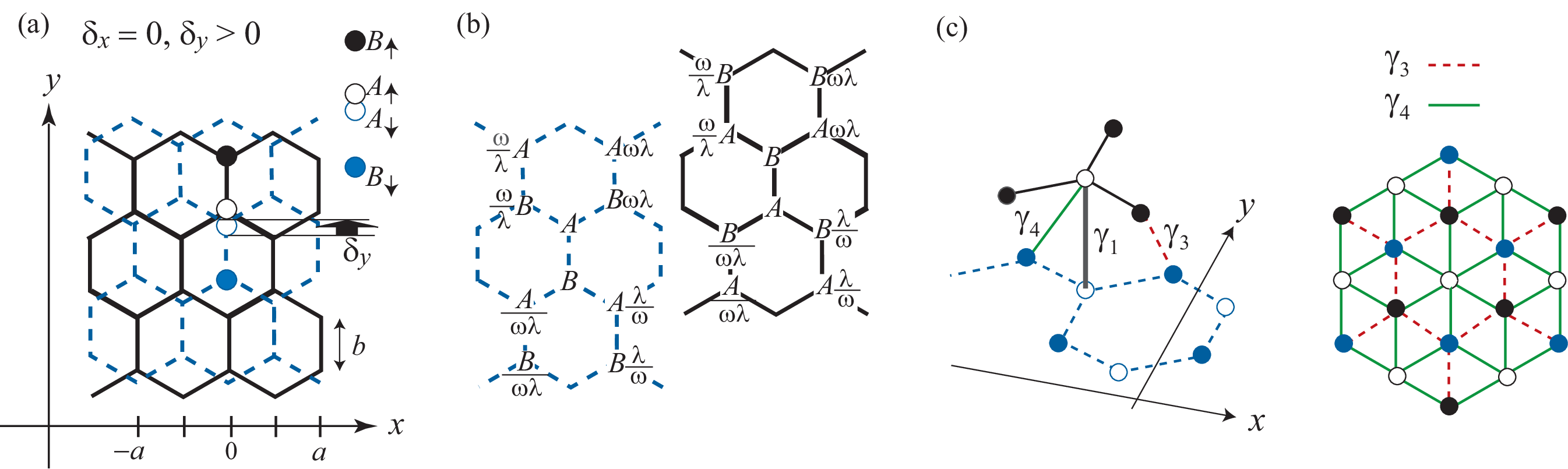}
\caption{
Atomic positions are labeled by integers $j_x$ and $j_y$ along the
$x$ and $y$ directions, respectively, with the constraint that
$j_x+j_y$ is even.
The vector $\vec{\delta}=(\delta_x,\delta_y)$ denotes the shear
displacement of the $\uparrow$ layer relative to the $\downarrow$
layer.
(a) Wave-function amplitudes
$(A_\downarrow,B_\downarrow,A_\uparrow,B_\uparrow)$ at
$j_x=j_y=0$ for $\delta_x=0$ and $b\gg\delta_y>0$, where
$b=a/\sqrt{3}$ is the in-plane bond length.
Dashed and solid lines denote the $\downarrow$ and $\uparrow$ layers,
respectively.
(b) Wave-function amplitudes in each layer for
$|j_x|\le1$ and $|j_y|\le1$.
The subscripts $\downarrow$ and $\uparrow$ are omitted for simplicity.
(c) Interlayer hopping parameters $\gamma_1$, $\gamma_3$, and
$\gamma_4$.
The left panel shows a schematic three-dimensional view, while the
right panel shows the corresponding projection onto the $(x,y)$
plane for $\delta_x=\delta_y=0$.
}
\end{center}
\end{figure*}

\section{Tight binding model and $\lambda$--$(1/\lambda^*)$ symmetry}

We employ a $\pi$-orbital tight-binding model with the standard notation for the hopping parameters
$(\gamma_0, \gamma_1, \gamma_3, \gamma_4)$.
The intralayer nearest-neighbor hopping parameter is denoted by $\gamma_0$, where
$\gamma_0=-|\gamma_0|=-3.12$~eV~\cite{Partoens}.
Figures~2(a) and 2(b) show the wave-function amplitudes
$(A_\downarrow,B_\downarrow,A_\uparrow,B_\uparrow)\lambda^{j_x}\omega^{j_y}$
with the Bloch factors
$\lambda=\exp(i k_xa/2)$ and
$\omega=\exp(i 3k_yb/2)$.
Here, $a=2.46$~\AA\  is the lattice constant, $b=a/\sqrt{3}$ is the bond length, and
$j_x$ and $j_y$ are integers satisfying the condition that $j_x+j_y$ is even.
As shown in Fig.~1(a), the $\uparrow$ layer, with a length of $(N-2)a/2$ ($N$ is an integer), is placed on top of the longer $\downarrow$ layer, and only the $\downarrow$ layer is connected to the source and drain electrodes.
 
 The interlayer hopping amplitudes are multiplied by exponential factors
$\exp(-\Delta r/l_d)$, where
$\Delta r$ denotes the change in the interatomic
distance induced by the shear displacement
$\vec{\delta}=(\delta_x,\delta_y)$ of the
$\uparrow$ layer relative to the $\downarrow$
layer.
These factors arise from the distance dependence
of the interlayer overlap between the $\pi$
orbitals.
Following Ref.~\cite{Lambin}, we adopt
$l_d=0.45$~\AA. Specifically, the hopping amplitudes are given by
\begin{equation}
\gamma_1=\gamma_1^{(0)}e^{\left (r_0-\tilde{r}_0 \right )/l_d} 
\label{gam-1}
\end{equation}
 for the  vertical $A_\downarrow$-$A_\uparrow$ hopping,
\begin{equation}
\gamma_{\;^3_4}^{(y)} =\gamma_{\;^3_4}^{(0)}e^{\left (r_1-\sqrt{\tilde{r}_1^2 \mp 2b \delta_y} \; \right )/l_d} 
\label{gam-y}
\end{equation}
for the skew hopping in the $y$  direction $(j_y \rightarrow j_y \pm 1)$, and 
\begin{equation}
\gamma_{\;^3_4}^{(+x)} =\gamma_{\;^3_4}^{(0)}e^{\left (r_1-\sqrt{\tilde{r}_1^2 \pm  b \delta_y+ a\delta_x} \; \right )/l_d} ,
\label{gam-x}
\end{equation}
\begin{equation}
\gamma_{\;^3_4}^{(-x)} =\gamma_{\;^3_4}^{(0)}e^{\left (r_1-\sqrt{\tilde{r}_1^2 \pm  b \delta_y- a\delta_x} \; \right )/l_d} 
\label{gam-x-2}
\end{equation}
 for the skew hopping in the $x$  direction $(j_x \rightarrow j_x \pm 1)$, 
 where
$r_0=3.35$~\AA\ is the interlayer distance,
$r_1=\sqrt{r_0^2+b^2}$,
$\tilde r_0=\sqrt{r_0^2+|\vec\delta|^2}$,
and
$\tilde r_1=\sqrt{r_1^2+|\vec\delta|^2}$.  
 Here, the positive direction of $\delta_y$ corresponds to the orientation
in which the angle between the $\gamma_1$ bond and the $\gamma_0$ bond
lying in the $yz$ plane is obtuse.
Following Ref.~\cite{Partoens}, the interlayer hopping amplitudes for $\vec{\delta}=0$ are taken to be $\gamma_1^{(0)}=0.377$~eV, $\gamma_3^{(0)}=0.29$~eV, and $\gamma_4^{(0)}=0.12$~eV.
 We consider Bernal-stacked bilayer graphene and employ a sublattice convention in which the vertical interlayer transfer $\gamma_1$ connects the same sublattice.
 The same convention is also used in Ref.~\cite{Partoens}.

  The tight-binding equation reads
\begin{equation}
  H\left(
\begin{array}{c}
\vec{\chi}_\downarrow\\
\vec{\chi}_\uparrow\\
\end{array}
\right)=E\left(
\begin{array}{c}
\vec{\chi}_\downarrow\\
\vec{\chi}_\uparrow\\
\end{array}
\right),
\label{TB-H}
  \end{equation}
  where
$\;^t\vec{\chi}_\downarrow=(A_\downarrow,B_\downarrow)$
and
$\;^t\vec{\chi}_\uparrow=(A_\uparrow,B_\uparrow)$
denote the layer-resolved wave functions.
The superscript $t$ denotes the transpose.
We write the Hamiltonian in block form
\begin{equation} 
H=
\left(
\begin{array}{cc}
h_+-\varepsilon {\bf 1}_2  & h_-' \\
h_+' & h_-+\varepsilon {\bf 1}_2
\end{array}
\right),
\label{def-H}
\end{equation}
where  the interlayer site energy difference is $2\varepsilon$,  and ${\bf 1}_n$ denotes the $n \times n$ identity matrix. 
The explicit forms of the blocks are
\begin{equation}
h_\pm=
|\gamma_0| \left(
\begin{array}{cc}
0  & (1 -\xi)\omega^{\pm 1}-1\\
(1 -\xi)\omega^{\mp 1}-1 & 0 
\end{array}
\right),
\label{h11}
\end{equation}
 \begin{equation}h_\pm'=
\left(
\begin{array}{cc}
\gamma_1  & f_4^{(\pm)}\\
 f_4^{(\pm)} & f_3^{(\pm)}
\end{array}
\right),
\end{equation}
\begin{equation}f_3^{(\pm)}\equiv \gamma_3^{(y)}\omega^{\pm 2} +\left[ \bar{\gamma}_3^{(x)} (\xi-1) \mp \widehat{\gamma}_3\eta \right]\omega^{\pm 1},
\label{f3}
\end{equation}
and 
\begin{equation}f_4^{(\pm)}\equiv \gamma_4^{(y)} + \left[ \bar{\gamma}_4^{(x)} (\xi-1) \mp \widehat \gamma_4\eta \right ]\omega^{\pm 1}.
\label{f4}
\end{equation}
The parameters $\xi$ and $\eta$ are determined by the Bloch factor $\lambda$
 as 
 \begin{equation}\xi\equiv \lambda+\frac{1}{\lambda} +1 
\label{xi-1}
\end{equation}
 and
\begin{equation}\eta \equiv \frac{1}{2}\left(\lambda-\frac{1}{\lambda}\right).
\label{eta}
\end{equation}
The definitions
\begin{equation}\bar{\gamma}_{3,4}^{(x)} \equiv \frac{1}{2}\left(\gamma_{3,4}^{(+x)}+\gamma_{3,4}^{(-x)}\right)\end{equation}
and 
\begin{equation}\widehat \gamma_{3,4} \equiv \gamma_{3,4}^{(+x)}-\gamma_{3,4}^{(-x)}\end{equation}
represent the average and the difference of $x$-shifted skew hoppings, respectively.
Note that $\widehat{\gamma}_{3,4}\,\delta_x < 0$, which will be important at the end of Sec. IV. 
In the effective theory, the wave number measured relative to the
$K$ and $K'$ points is defined as
$\tilde{k}_x = k_x \pm 4\pi/(3a)$.
  Here, $k_x$ and $\tilde{k}_x$ are \emph{not} necessarily real numbers.
  Appendix~A shows that $\tilde{k}_x$ is related to Eq.~(\ref{xi-1}) via $\tilde{k}_xa=\pm 2\xi/\sqrt{3}$~\cite{pseudo-spin-1,pseudo-spin-2,note-1}.
 Since we consider the case where $|\varepsilon| \ll |\gamma_0| $ and $|E| \ll |\gamma_0|$, $\xi$ and $\tilde{k}_x$ are close to zero.

The identity $\det({}^t C^*) = (\det C)^*$ holds for an arbitrary matrix $C$,
where the superscript $*$ denotes complex conjugation.
Therefore, the secular equation
\begin{equation}
\det (E{\bf 1}_4 - H) = 0
\label{eq:3}
\end{equation}
is equivalent to
\begin{equation}
\det (E{\bf 1}_4 - {}^t\!H^*) = 0,
\label{eq:4}
\end{equation}
where the matrix $H$ is defined in Eq.~(\ref{def-H}).
 Within the energy gap, $|\lambda| \neq 1$, and therefore $\xi^* \neq \xi$, 
$(f_3^{\pm})^* \neq f_3^{\mp}$ and
$(f_4^{\pm})^* \neq f_4^{\mp}$.
Hence, $H$ is \emph{not} Hermitian
($H \neq {}^t\!H^*$).
Replacing $H$ with ${}^t\!H^*$ is equivalent to replacing $\lambda$ with $1/\lambda^*$.
Thus, if $\lambda$ is an allowed Bloch factor, so is $1/\lambda^*$ for the same $k_y$.
We refer to this as the $\lambda$--$(1/\lambda^{*})$ symmetry, which originates from the translational periodicity along the $x$ axis.
This symmetry significantly constrains the allowed solutions for $\lambda$.
First, it is sufficient to consider only $|\lambda| < 1$, since the partner $1/\lambda^*$ has magnitude greater than unity.
Second, for fixed $k_y$, the symmetry can be expressed as $(\lambda, k_y)$--$(1/\lambda^{*}, k_y)$.
Note that $E$, $\varepsilon$, and $k_y$ are identical in Eqs.~(\ref{eq:3}) and (\ref{eq:4}).
Time-reversal symmetry ensures $(\lambda, k_y)$--$(\lambda^{*}, -k_y)$, but not $(\lambda, k_y)$--$(\lambda^{*}, k_y)$.
However, when $\delta_x = 0$, the system also possesses $(\lambda, k_y)$--$(1/\lambda, k_y)$ symmetry.
Combining these relations yields $(\lambda, k_y)$--$(\lambda^{*}, k_y)$ symmetry and $(\lambda, k_y)$--$(\lambda, -k_y)$ symmetry.
Consequently, for $\delta_x = 0$, it is sufficient to present $(k_x^{\rm r}, k_x^{\rm i})$ only for the $K$ valley ($k_x^{\rm r} a/2 \simeq -2\pi/3$), even when $k_y \neq 0$,
whereas both valleys are required when $\delta_x k_y \neq 0$.
Table~I summarizes these symmetry transformations of the Bloch factors.

\begin{table}
\caption{\label{tab:table3}
Transformation of the Bloch factors
$(\lambda, \omega)=\left(\exp(ik_xa/2),\,\exp(i3k_yb/2)\right)$
and the wave numbers
$(k_x^{\rm r}, k_x^{\rm i}, k_y)$
under the symmetry operations $P$.
We use the notation
$A^{\mathrm{r}}=\mathrm{Re}(A)$ and
$A^{\mathrm{i}}=\mathrm{Im}(A)$
for the real and imaginary parts of a complex number $A$, respectively.
The wave number
$k_x=k_x^{\rm r}+ik_x^{\rm i}$
is complex with
$k_x^{\rm i}\neq0$,
whereas $k_y$ is real.
}
{
\renewcommand{\arraystretch}{1.3}
\begin{tabular}{|c|c|c|}
\hline
operation $P$ & $P(\lambda, \omega)$ & $P(k_x^{\rm r}, k_x^{\rm i}, k_y)$ \\ \hline
$H \to {}^t\!H^*$ & $(1/\lambda^*, \omega)$ & $(k_x^{\rm r}, -k_x^{\rm i}, k_y)$ \\ \hline
 $H \to H^*$ time-reversal & $(\lambda^*, \omega^*)$ & $(-k_x^{\rm r}, k_x^{\rm i}, -k_y)$ \\ \hline
$x \to -x$ inversion & $(1/\lambda, \omega)$ & $(-k_x^{\rm r}, -k_x^{\rm i}, k_y)$ \\ \hline
\end{tabular}
}
\end{table}

\section{$\widetilde{\Delta}\xi$ approximation}

Equations~(\ref{xi-1}) and (\ref{eta}) can be rewritten as
\begin{equation}
\lambda^{(\sigma)}
=\frac{\xi-1}{2}
+ \sigma \sqrt{\left(\frac{\xi-1}{2}\right)^2-1}
\label{lambda-xi}
\end{equation}
and
\begin{equation}
\eta
=\sigma \sqrt{\left(\frac{\xi-1}{2}\right)^2-1},
\label{eta-xi}
\end{equation}
where $\sigma=\pm$.
When $k_y\delta_x\neq 0$, the $\lambda$--$(1/\lambda)$ symmetry is absent, and hence
$\lambda^{(+)}\lambda^{(-)}\neq 1$.
This is because the secular equation~(\ref{eq:3}) depends not only on $\xi$ but also on $\eta$,
so that the solution $\xi$ satisfying Eq.~(\ref{eq:3}) depends on $\sigma$ as well.
In what follows, we use the notation
$A^{\mathrm{r}}=\mathrm{Re}(A)$ and
$A^{\mathrm{i}}=\mathrm{Im}(A)$
for the real and imaginary parts of a complex number $A$, respectively,
and define $\mathrm{sgn}(B)=B/|B|$ for a real number $B$.
Equation~(\ref{lambda-xi}) can be approximated as
\begin{equation}
\lambda^{(\sigma)} \simeq
\exp \left(
-\sigma\frac{|\xi^{\mathrm{i}}|}{\sqrt{3}}
+i\sigma\,\mathrm{sgn}(\xi^{\mathrm{i}})
\left(\frac{\xi^{\mathrm{r}}}{\sqrt{3}}-\frac{2}{3}\pi\right) \right)
\label{lambda-xir-xii}
\end{equation}
for a nonzero $\xi^{\mathrm{i}}$, and
\begin{equation}
\lambda^{(\sigma)} \simeq
\exp \left[ \sigma i\left(\frac{\xi}{\sqrt{3}}-\frac{2}{3}\pi\right) \right]
\label{lambda-xir}
\end{equation}
for $\xi^{\mathrm{i}}=0$.
See Appendix~A for the derivation.
When $|\lambda|=1$, Eq.~(\ref{lambda-xir}) shows that $\sigma$ serves as the valley index.

\begin{table}
\caption{
Correspondence between the $K$ and $K'$ valleys and the sign indices
$(\sigma,\tau,l)$ for $k_y=0$ and $\gamma_3=\gamma_4=0$.
Here, the $K$ and $K'$ valleys correspond to
${\rm Im}(\lambda^2)>0$ and ${\rm Im}(\lambda^2)<0$,
respectively, where
$\lambda^2=\exp(ik_xa)$.
Region (i) includes both the energy-gap region,
$|E|<E_g$, and the high-energy region,
$\sqrt{\varepsilon^2+\gamma_1^2}<|E|$.
Region (ii) corresponds to the reduced-channel region,
where $iq=|q|$ and $p-|q|<0$.
}
{
\renewcommand{\arraystretch}{1.2}
\begin{tabular}{c|c|c|c}
\multicolumn{4}{l}{ (i) $|E|<|\varepsilon|$, $\sqrt{\varepsilon^2+\gamma_1^2}< |E|$} \\ \hline
  &$|\lambda|=1$ & $|\lambda| <1$ & $|\lambda| >1$ \\ \hline
        & arbitrary $\tau$ and $l$ & $\sigma=+ $ & $\sigma=-$ \\ \hline
  $K$ & $\sigma=+$ & $\tau l=-$  & $\tau l=+$  \\ \hline
  $K'$ & $\sigma=-$ & $\tau l=+$  & $\tau l=-$  \\ \hline
\end{tabular}
}

\vspace{.5cm}
{
\renewcommand{\arraystretch}{1.2}
\begin{tabular}{c|c|c|c}
\multicolumn{4}{l}{ (ii) $|\varepsilon| <|E|< \sqrt{\varepsilon^2+\gamma_1^2}$} \\ \hline
  &$|\lambda|=1$ & $|\lambda| <1$ & $|\lambda| >1$ \\ \hline
  & $l=+$  &\multicolumn{2}{c}{ $l=-$} \\ \hline
  & arbitrary $\tau$  & $\sigma=+$ & $\sigma=-$ \\ \hline
  $K$ & $\sigma=+$ & $\tau =-$  & $\tau =+$  \\ \hline
  $K'$ & $\sigma=-$ & $\tau =+$  & $\tau =-$  \\ \hline
\end{tabular}
}
\end{table}

We introduce 
\begin{equation}
X=\frac{1}{\gamma_0^4}\det(E {\bf 1}_4-H)
\label{def-X}
\end{equation}
 and regard the equation $X=0$ as an equation for determining $\xi$.
When $\gamma_3=\gamma_4=0$, the exact solution for $\xi$  is given by
\begin{equation} 
\xi_{\tau,l}^{(0)}=1-c -\tau \sqrt{p-s^2+ilq}
\label{xi-zero}
 \end{equation}
where $\tau=\pm$, $l=\pm$, 
 $s\equiv \sin\left(\frac{3}{2}k_y b\right), c \equiv \cos\left(\frac{3}{2}k_y b\right), p\equiv (E^2+\varepsilon^2)/\gamma_0^2$, 
\begin{equation}
q \equiv \left\{
\begin{array}{ccc}
\gamma_0^{-2} \sqrt{ (4\varepsilon^2+\gamma_1^2)(E_g^2-E^2)}& \cdots & |E| < E_g\\
-i\gamma_0^{-2} \sqrt{(4\varepsilon^2+\gamma_1^2)(E^2-E_g^2)} & \cdots & |E|  > E_g
\end{array} 
\right.,
\label{def-q}
\end{equation} 
with the half-gap width of the $\gamma_1$-only tight-binding model, 
\begin{equation}
E_g=\frac{|\varepsilon|\gamma_1}{\sqrt{4\varepsilon^2+\gamma_1^2}}.
\end{equation}
Table~II shows correspondences between the valleys and the sign indices
for $k_y=0$ and $\gamma_3=\gamma_4=0$.
The correspondences listed in this table are approximately preserved
for small $k_y$, $\gamma_3$, and $\gamma_4$.
Owing to the $\lambda$--$(1/\lambda^{*})$ symmetry, the discussion can be restricted to
the case $|\lambda| \leq 1$.
When $|\lambda| \neq 1$, we set $\sigma=+$ and abbreviate
$\lambda^{(+)}$ as $\lambda$.

Throughout this paper, we consider a MBMGJ with the current along the $x$ direction; in this case, propagating states in the monolayer leads require $|s|<|E+\varepsilon|/|\gamma_0|$.
Since $E+\varepsilon \ll |\gamma_0|$, $\omega$ becomes close to unity.
Therefore, when the displacement is also small, 
Eqs.~(\ref{f3}) and (\ref{f4}) reduce approximately to
$\gamma_3^{(0)}\xi$
and
$\gamma_4^{(0)}\xi$,
respectively.
Although $\gamma_{3}$ and $\gamma_{4}$ have magnitudes comparable to $\gamma_{1}$, the smallness of $\xi$ allows us to treat $f_{3}$ and $f_{4}$ as perturbations.
Neglecting the second and higher-order terms in $f_{3}$ and $f_{4}$, the function $X$ can be approximated as
\begin{equation}
X =  \prod_{\tau=\pm} \prod_{l=\pm}\left(\xi-\xi_{\tau,l}^{(0)}\right)+\frac{\gamma_1}{\gamma_0^2}\left(D_{\eta} +\sum_{n=0}^3 D_n \xi^n \right),
\label{X}
\end{equation}
 where
 \begin{equation}
 D_0=4(c-1)\left[ \bar{\gamma}_3^{(x)} -c\gamma_3^{(y)} -\frac{E}{|\gamma_0|}(\bar{\gamma}_4^{(x)}+\gamma_4^{(y)})\right],
 \end{equation}
 \begin{eqnarray}
D_1 &=&4(1-c)(\gamma_3^{(y)}+2\bar{\gamma}_3^{(x)})\nonumber \\  
 && + 4\frac{E}{|\gamma_0|}\left((c-2)\bar{\gamma}_4^{(x)}+c\gamma_4^{(y)}\right),
\end{eqnarray}
 \begin{equation}
D_2= (6c-4)\bar{\gamma}_3^{(x)}-2\gamma_3^{(y)}+4\frac{E}{|\gamma_0|}\bar{\gamma}_4^{(x)},
\end{equation}
 \begin{equation}
D_3=-2c\bar{\gamma}_3^{(x)},
\end{equation}
and
\begin{equation}
D_\eta=2is\eta \left[ (2\xi-\xi^2)\widehat\gamma_3-2\frac{E}{|\gamma_0|}\widehat\gamma_4\right].
\label{D-eta}
\end{equation}
Owing to the approximation $\eta \simeq i \tau l \sqrt{3}/2$,
which is justified by Eq.~(\ref{lambda-xir-xii}) and $\sigma=+$,
the equation $X=0$ can be approximately regarded as a quartic equation in $\xi$,
even when $\delta_x k_y \neq 0$.
This allows us to apply the Durand--Kerner method~\cite{Durand,Kerner} to obtain the shift
$\widetilde{\Delta}\xi \equiv \xi - \xi^{(0)}$
 caused by the skew hopping, yielding
\begin{eqnarray}
 \widetilde{\Delta}\xi_{\tau,l} &= & \left. \frac{ -X }{ \left(\xi-\xi_{-\tau,l}^{(0)}\right)\left(\xi
-\xi_{\tau,-l}^{(0)}\right)\left(\xi_-\xi_{-\tau,-l}^{(0)}\right)}\right|_{\xi=\xi_{\tau,l}^{(0)}}
\\
&= &\left. \frac{ -\gamma_1 D_\eta-\gamma_1 \sum_{n=0}^3 D_n\xi^n }{-4ilq\tau\gamma_0^2\sqrt{p-s^2+ilq }}\right|_{\xi=\xi_{\tau,l}^{(0)}}.
\label{DK-eq}
\end{eqnarray}
We refer to the resulting approximation as the
$\widetilde{\Delta}\xi$ approximation.
The change induced by $\vec{\delta}$ is given by
\begin{equation}
\Delta\xi_{\tau,l} \equiv \left. \widetilde{\Delta}\xi_{\tau,l} \right|_{\vec{\delta} \neq 0}  -\left. \widetilde{\Delta}\xi_{\tau,l} \right|_{\vec{\delta} = 0} .
\end{equation}

\section{Dispersion lines}
\begin{figure}
\begin{center}
\includegraphics[width=\linewidth]{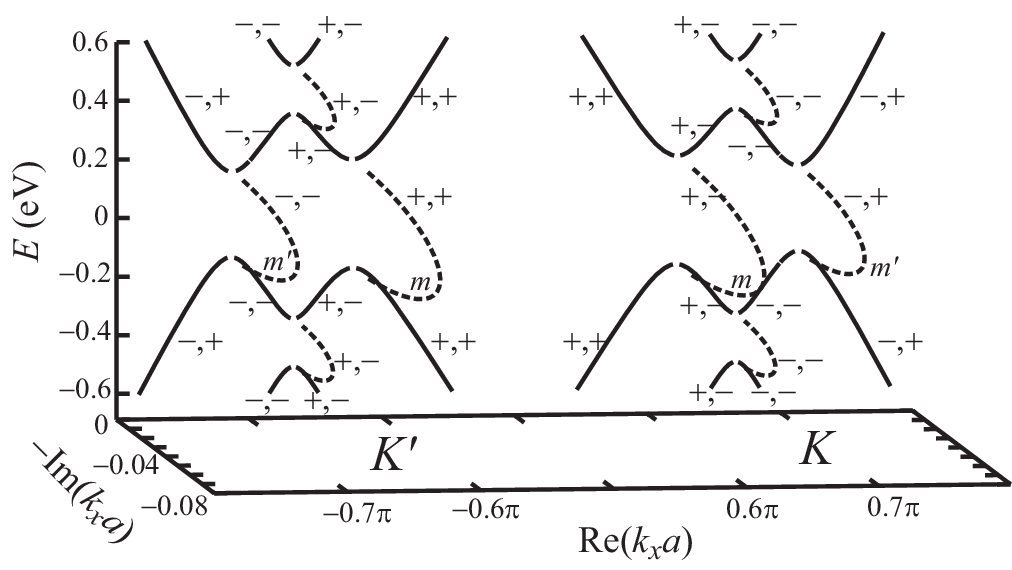}
\caption{
The relation between $(k_x^{\mathrm{r}}a,-k_x^{\mathrm{i}}a)$ and the
energy $E$ for $\varepsilon=0.35~\mathrm{eV}$, $k_y=0$, and
$\vec{\delta}=0$.
Here, $k_x^{\mathrm{r}}=\mathrm{Re}(k_x)$ and
$k_x^{\mathrm{i}}=\mathrm{Im}(k_x)$.
Solid and dashed lines represent the RDLs and CDLs, respectively.
Because the data were calculated with an energy interval of
0.004~eV, the CDL and RDL appear disconnected.
In fact, they are connected at the extrema of the RDL.
Each DL is labeled by $(\tau,l)$.
See the text and Table~II for the definitions of
$(\tau,l)$, $m$, and $m'$.
}
\end{center}
\end{figure}

Figure~3 shows the exact dispersion lines (DLs), representing the relation between $(k_x^{\mathrm{r}} a, -k_x^{\mathrm{i}} a)$ and the energy $E$, for $\varepsilon = 0.35~\mathrm{eV}$, $k_y = 0$, and $\vec{\delta} = 0$.
They are obtained by numerical diagonalization of the transfer matrix, as described in the Appendix~B.
Each DL is labeled by the index $(\tau,l)$.
When $k_x^{\mathrm{i}} \neq 0$, only the CDLs corresponding to $|\lambda| < 1$
are indicated.
In the gap region, no RDLs exist, whereas CDLs remain.
CDLs connect local extrema (minima and maxima) belonging to different RDLs.
Among the zero-energy points of the CDLs in the gap,
we denote by $m$ the point satisfying $|k_x^{\mathrm{r}} a| < 2\pi/3$,
and by $m'$ the point satisfying $|k_x^{\mathrm{r}} a| > 2\pi/3$.
The CDLs passing through these points are referred to as
the $m$DL and $m'$DL, respectively.
When $\gamma_3 = \gamma_4 = 0$, the $m$DL and $m'$DL connect to an RDL
at a common energy.
However, skew hopping shifts the connection energies of the $m$DL
and $m'$DL relative to each other.
To emphasize this shift, we choose a relatively large value of
$\varepsilon$ (0.35 eV) in Fig.~3.
In the following, we instead use
$\varepsilon = 0.15~\mathrm{eV}$.
This value is close to $\gamma_1/2$, corresponding to the experimentally relevant gap~\cite{gra-Gap-experiment-1,gra-Gap-experiment-2,gra-Gap-experiment-3,gra-Gap-experiment-4}.
For the effects of the shear displacement,
the role of the skew hoppings becomes more significant than that of
$\gamma_1$.
This is because the change in skew hopping due to the displacement
$\vec{\delta}$ appears only at second order in $|\vec{\delta}|$ in
Eq.~(\ref{gam-1}),
whereas it enters linearly in $\delta_y$ and $\delta_x$ in
Eqs.~(\ref{gam-y}),  (\ref{gam-x}), and~(\ref{gam-x-2}).

Figure~4 shows the DLs in the $K$ valley for $\vec{\delta}=0$ at
(a) $k_y=0$,
(b) $3k_yb=0.03\pi$, and
(c) $3k_yb=0.032\pi$.
The solid lines represent exact results, as in Fig.~3.
The circles show results obtained within the $\widetilde{\Delta}\xi$ approximation, evaluated at $E=-0.5, -0.49, \ldots, 0.5$~eV with an energy spacing of $0.01$~eV.
The RDLs are labeled in ascending order of energy as
$E_1$, $E_2$, $E_3$, and $E_4$.
For the extrema of $E_2$ and $E_3$, those connected to the $m$DL
are labeled $2$ and $3$,
those connected to the $m'$DL are labeled $2'$ and $3'$,
and those connected to $E_1$ or $E_4$ are labeled $2''$ and $3''$.
As $k_y$ increases, $3'$ and $2'$ approach $3''$ and $2''$, respectively,
and they eventually merge and disappear.
In Fig.~4(b), $3'$ and $3''$ vanish, and the corresponding CDLs become detached from $E_3$.
As a result, two CDLs merge and the $m'$DL connects to  $E_4$.
As $k_y$ increases further, $2'$ and $2''$ also disappear,
and a CDL connecting $E_1$ and $E_4$ emerges.
The CDL generated by the merger retains a memory of its original shape
and subsequently drifts.
As is clear from the analytical expressions,
when $q$ is close to zero, i.e., near the gap edge of the
$\gamma_1$-only tight-binding model,
the $\widetilde{\Delta}\xi$ approximation diverges and becomes invalid.
Accordingly, circles are not shown when $|q|^2 < 10^{-5}$.
Away from this region, the
$\widetilde{\Delta}\xi$
approximation accurately reproduces both the CDLs and the RDLs.

\begin{figure}
\begin{center}
\includegraphics[width=\linewidth]{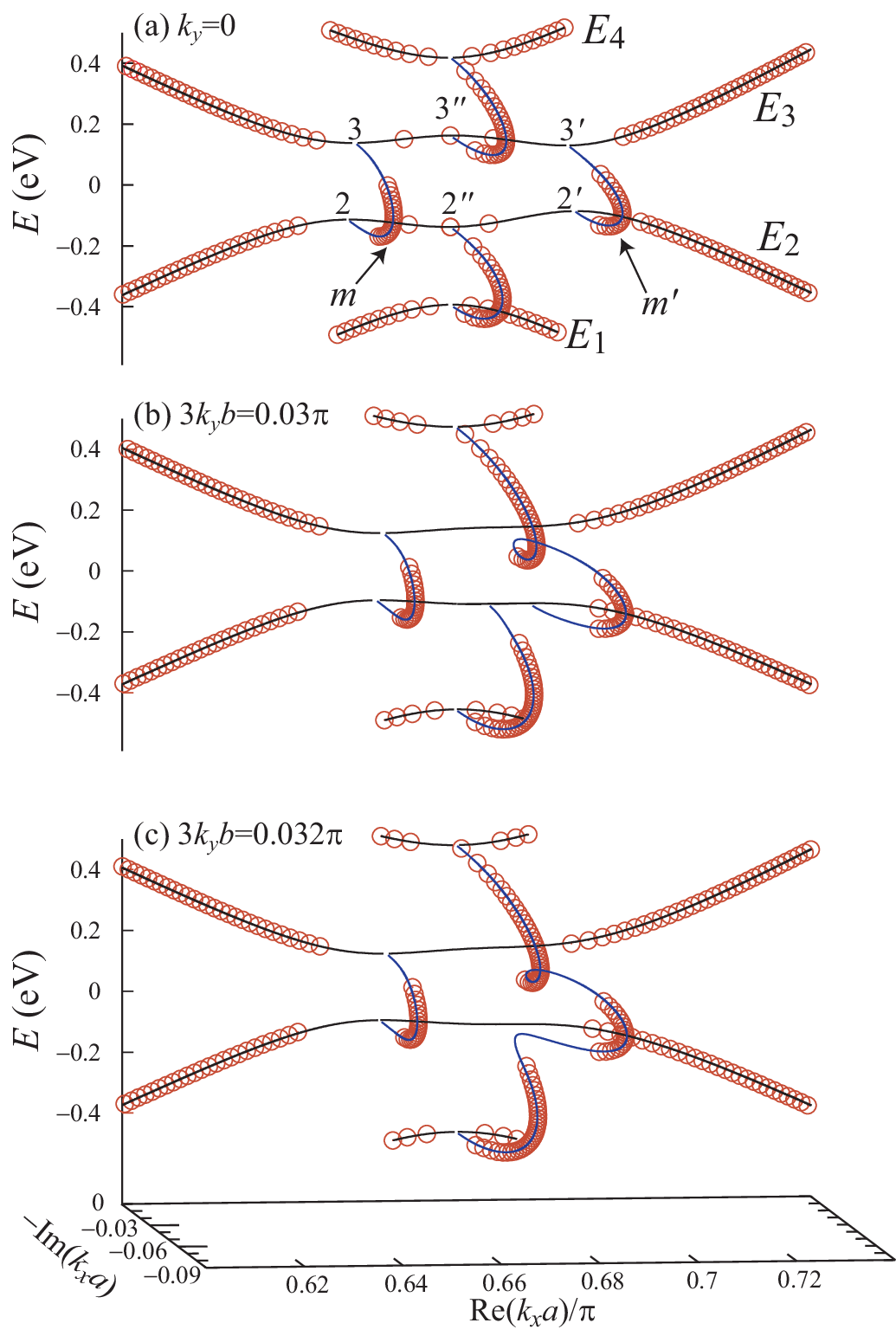}
\caption{
The DLs in the $K$ valley for $\vec{\delta}=0$ at
(a) $k_y=0$,
(b) $3k_yb=0.03\pi$, and
(c) $3k_yb=0.032\pi$, for $\varepsilon=0.15$~eV.
Solid lines represent exact results, while circles correspond to the
$\widetilde{\Delta}\xi$ approximation.
For the $\widetilde{\Delta}\xi$ approximation, calculations are
performed at $E=-0.5,-0.49,\ldots,0.5$~eV with an energy spacing of
0.01~eV, except in the region where $|q|^2<10^{-5}$.
The RDLs are labeled in ascending order of energy as
$E_1$, $E_2$, $E_3$, and $E_4$.
See the text for the definitions of the points
$m$, $m'$, $2$, $2'$, $2''$, $3$, $3'$, and $3''$.
}
\end{center}
\end{figure}

\begin{table*}
\caption{\label{tab:table1}
Complex wave numbers $k_x$ at $E=0$ for different shear displacements
$\vec{\delta}=(\delta_x,\delta_y)$~\AA, including the case
$\vec{\delta}=(0,0)$,
evaluated within the $\widetilde{\Delta}\xi$ approximation and presented in the format
$(|k_x^{\mathrm{r}}|a/\pi - 0.66,\; -k_x^{\mathrm{i}}a)\times10^{4}$.
See the text for the definitions of $m$ and $m'$.
}
\begin{ruledtabular}
\begin{tabular}{lccccccc}
$(\delta_x,\delta_y)$~\AA & valley &\multicolumn{2}{c}{$k_y=0$}&\multicolumn{2}{c}{$3k_yb=0.02\pi$}&\multicolumn{2}{c}{$3k_yb=0.04\pi$}\\
 & & $m$ & $m'$ & $m$ & $m'$ & $m$ & $m'$ \\
$(0,0)$& $K$ 
 & ($-$154,   $-$549) 
  &
(326,   $-$476)
&
($-$134,   $-$546)
&
(310,   $-$560)
&
($-$81,   $-$568)
&
(271,   $-$844)
\\ \hline
$(0,0.1)$& $K$ 
&
($-$177,   $-$449)
&
(345,   $-$380)
&
($-$162,   $-$471)
&
(334,   $-$484)
&
($-$130,   $-$552)
&
(317,   $-$811)
\\ \hline
$(0,-0.1)$& $K$ 
 &
($-$132,   $-$642)
&
(307,   $-$565)
&
($-$107,   $-$614)
&
(286,   $-$630)
&
($-$33,   $-$581)
&
(227,   $-$875)
\\ \hline
$(0.1,0)$& $K$ 
&
($-$153,   $-$551)
&
(325,   $-$477)
&
($-$132,   $-$639)
&
(308,   $-$472)
&
($-$76,   $-$750)
&
(267,   $-$664)
\\ \hline
$(0.1,0)$& $K'$ 
 &
($-$153,   $-$551)
&
(325,   $-$477)
&
($-$134,   $-$453)
&
(310,   $-$649)
&
($-$82,   $-$384)
&
(273,   $-$1024)
\\ \hline
\end{tabular}
\end{ruledtabular}
\end{table*}

\begin{table*}
\caption{\label{tab:table2}
Changes in the complex wave numbers $k_x$ at $E=0$ induced by the shear
displacement $\vec{\delta}$, evaluated within the
$\widetilde{\Delta}\xi$ approximation and presented in the same format
as in Table~III.
The values represent
$k_x(\vec{\delta}\neq0)-k_x(\vec{\delta}=0)$.
}
\begin{ruledtabular}
\begin{tabular}{lccccccc}
$(\delta_x,\delta_y)$~\AA &valley &\multicolumn{2}{c}{$k_y=0$}&\multicolumn{2}{c}{$3k_yb=0.02\pi$}&\multicolumn{2}{c}{$3k_yb=0.04\pi$}\\
 &  &$m$ & $m'$ & $m$ & $m'$ & $m$ & $m'$ \\
$(0,0.1)$& $K$ 
&
($-$23,   100)
&
(19,   95)
&
($-$28,   75)
&
(24,   76)
&
($-$50,   15)
&
(46,   34)
\\ \hline
$(0,-0.1)$& $K$  &
(22,   $-$93)
&
($-$19,   $-$89)
&
(27,   $-$69)
&
($-$24,   $-$70)
&
(48,   $-$13)
&
($-$45,   $-$31)
\\ \hline
$(0.1,0)$ &$K$ 
&
(1,   $-$1)
&
($-$1,   $-$1)
&
(2,   $-$94)
&
($-$2,   88)
&
(4,   $-$182)
&
($-$5,   180)
\\ \hline
$(0.1,0)$ &$K'$   &
(1,   $-$1)
&
($-$1,   $-$1)
&
(0,   92)
&
(0,   $-$89)
&
($-$1,   183)
&
(2,   $-$179)
\\ \hline
\end{tabular}
\end{ruledtabular}
\end{table*}

Table~III lists the values of
$(|k_x^{\mathrm{r}}|a/\pi - 0.66,\; -k_x^{\mathrm{i}}a)\times 10^{4}$
at the $m$ and $m'$ points,
evaluated within the $\widetilde{\Delta}\xi$ approximation,
while Table~IV summarizes the corresponding changes induced by the displacement
$\vec{\delta}$.
The corresponding CDLs are shown in Appendix~C,
 demonstrating that the $\widetilde{\Delta}\xi$ approximation remains valid
 even when $\vec{\delta} \neq 0$.
A positive $\delta_y$ shifts $|\lambda|$ toward unity, 
suggesting an enhancement of the conductance $G$ 
of the MBMGJ. 
Conversely, a negative $\delta_y$ drives $|\lambda|$ away from unity 
and is expected to reduce $G$.
The changes in $|\lambda|$ induced by $\delta_x$
have opposite signs at $m$ and $m'$ and tend to cancel out.
Since the sign of $\delta_x s$ is fixed,
this cancellation is not due to a sign reversal of $\delta_x s$ and occurs within a single valley.
 This cancellation suggests that the effect of
$\delta_x$
on the conductance
of a MBMGJ is weaker than that of
$\delta_y$.

To provide a physical interpretation of Table~IV, we further simplify Eq.~(\ref{DK-eq}) under the following conditions.
(i) As discussed before Eq.~(\ref{X}),  we consider only values of $s$ close to zero.
(ii) Since $E$ is close to zero, the term $\gamma_4$ is neglected.
(iii) Because $\bar{\gamma}_3^{(x)}$ is insensitive to the shear displacement $\vec{\delta}$, we neglect $D_3$.
Condition (i) implies $-\tau\sqrt{p-s^2+ilq} \simeq \xi^{(0)}_{\tau,l}$.
Under these conditions, we obtain
\begin{equation}
\Delta \xi_{\tau,l} \simeq -i \frac{\sqrt{3}\gamma_1}{2q\gamma_0^2} s\,\tau\,\widehat{\gamma}_3
 - i l \theta \, \xi_{\tau,l}^{(0)} ,
 \label{simple-dxi}
\end{equation}
where
\begin{equation}
\theta \equiv \frac{\gamma_1}{2|q|\gamma_0^2} \left( \gamma_3^{(y)} - \gamma_3^{(0)} \right).
\label{theta}
\end{equation}
Although we take $q$ as a positive real number in this section,  $|q|$ is used in Eq.~(\ref{theta}) to allow for the case discussed later, where $q$ becomes purely imaginary.
Using $c-1 \simeq -s^2/2$, we neglect the terms $D_0$ and $D_1$, which are of second order in $s$, while retaining $D_\eta$, which is linear in $s$.
Furthermore, since $|\xi| \ll 1$, only terms linear in $\xi$ are retained in Eq.~(\ref{D-eta}).
From Eq.~(\ref{simple-dxi}),  we derive 
\begin{equation}
\Delta\left |\xi_{\tau,l}^{\mathrm{i}}\right| =  \frac{\sqrt{3}\gamma_1}{2q\gamma_0^2} sl\widehat\gamma_3
-\theta \left|\mathrm{Re}
(\xi_{\tau,l}^{(0)}) \right|,
\label{app-Im-Dxi}
\end{equation}
and 
\begin{equation}
\Delta\left |\xi_{\tau,l}^{\mathrm{r}}\right| = \theta \left|\mathrm{Im}(\xi_{\tau,l}^{(0)}) \right| .
\label{app-Re-Dxi}
\end{equation}
Here, we use the relations $|\xi^{\mathrm{i}}| = \mathrm{sgn}(\xi^{\mathrm{i}})\,\xi^{\mathrm{i}}=-\tau l \xi^{\mathrm{i}}$
 and $|\xi^{\mathrm{r}}| = \mathrm{sgn}(\xi^{\mathrm{r}})\,\xi^{\mathrm{r}}=-\tau \xi^{\mathrm{r}}$.
 This assumes that the perturbation $f_3$ does not alter the signs of real and imaginary parts of $\xi_{\tau,l}$.
 $\widehat{\gamma}_3$ and $\theta$ respond negatively to $\delta_x$ and
positively to $\delta_y$, respectively.
Eqs.~(\ref{app-Im-Dxi}) and~(\ref{app-Re-Dxi}),
together with
$|\xi^{\mathrm{r}}/\sqrt{3}| \simeq |k_x^{\mathrm{r}}a\mp 2\pi/3|$
and the signs of the responses of
$\widehat{\gamma}_3$ and $\theta$ to
$\delta_x$ and $\delta_y$,
are consistent with Table~IV.
  As shown in Fig.~3, the $m$ and $m'$ points correspond to
$l=-$ and $l=+$ in the $K$ valley, and to
$l=+$ and $l=-$ in the $K'$ valley, respectively.
  It should be emphasized that Eq.~(\ref{simple-dxi}) is introduced solely for the purpose of providing a physical interpretation of the results.
All numerical results within the $\widetilde{\Delta}\xi$ approximation are obtained directly from the original Eq.~(\ref{DK-eq}).

\section{Conductance of monolayer/bilayer/monolayer graphene junctions }
We consider MBMGJs ($\downarrow$--$\downarrow\uparrow$--$\downarrow$ graphene junctions), in which the barrier thickness is determined by the length of the central $\uparrow$ layer, $(N-2)a/2$.
 The definition of $N$ follows Ref.~\cite{Tamura-2025}, and Fig.~1(b) corresponds to the case $N=6$.
 Following the Bloch-factor convention introduced in Secs. III and IV, the transport direction is taken along the $x$ axis, and the interfaces between the monolayer and bilayer regions are assumed to be ideal armchair boundaries. 
Appendix~B shows that the monolayer--bilayer armchair boundary is rigorously treated in the exact calculation of the transmission probability.
\begin{figure}
\begin{center}
\includegraphics[width=\linewidth]{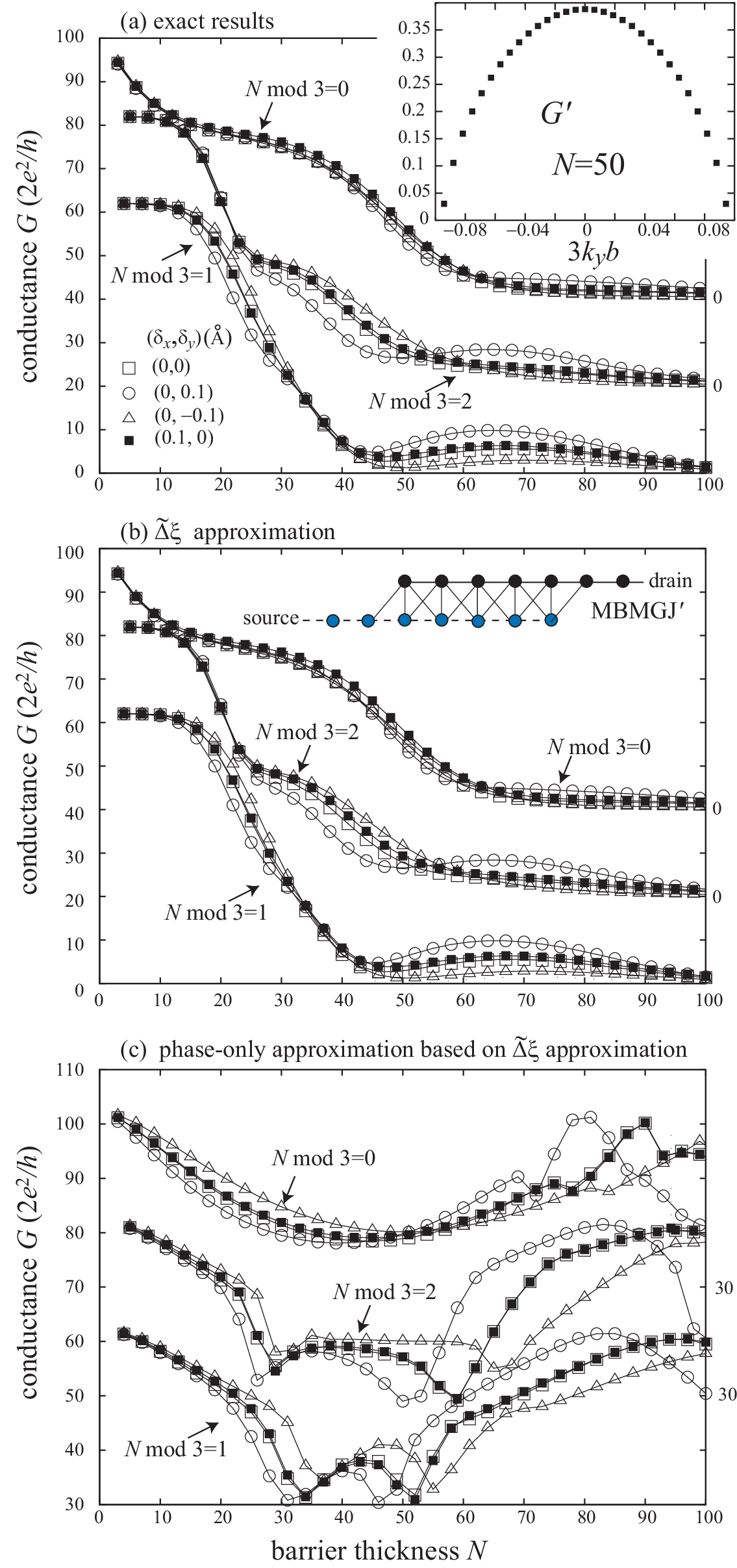}
\caption{
The zero-energy conductance $G$ of MBMGJs, calculated from the
Landauer formula with periodic boundary conditions in the $y$
direction for a transverse width of $3000b$, for
$\varepsilon=0.15$~eV.
For $N \bmod 3 = 2$ and $N \bmod 3 = 0$, the data are vertically
shifted by 20 and 40, respectively, in units of
$G_0=2e^2/h$.
Here, $N \bmod 3$ denotes the remainder when $N$ is divided by 3.
The inset in (a) shows the $k_y$-resolved conductance
$G'(k_y)$ for $E=0$, $\vec{\delta}=(0.1,0)$~\AA,
and $N=50$, evaluated for
$3k_yb=0,\pm\frac{\pi}{500},
\pm\frac{2\pi}{500},\ldots,
\pm\frac{15\pi}{500}$.
The inset in (b) shows the side view of another
monolayer/bilayer/monolayer
($\downarrow$/$\downarrow\uparrow$/$\uparrow$)
graphene junction, denoted by MBMGJ$'$,
discussed in Ref.~\cite{Tamura-2025}.
}
\end{center}
\label{Fig5}
\end{figure}
Figure~5(a) shows the zero-energy conductance $G$ of MBMGJs
calculated using the Landauer formula and plotted as a function of $N$  in units of $G_0=2e^2/h$~\cite{Datta,T-Ando,Landauer}.
In the exact numerical calculation shown in Fig.~5(a), 
the Bloch factor
$\tilde{\lambda}$
is obtained by numerically diagonalizing the transfer matrix.
Within the single-particle approximation,
the scattering-matrix approach used in the present
exact calculation is equivalent to the
Green's-function method~\cite{Datta}.
 Figures ~5(b) and ~5(c) show the results obtained by replacing $\tilde{\lambda}$ in the same numerical code with $\lambda$ and $\lambda/|\lambda|$, respectively. 
Here, $\lambda$ denotes the approximate quantity $\exp(i k_x a/2)$ obtained from Eq.~(\ref{DK-eq}). The details of the calculation are provided in the Appendix~B~\cite{Tamura-exact}. 

The conductance is given by
\begin{equation}
G=\sum_nG'\left(\frac{2\pi}{L_y}n\right),
\label{cond-G}
\end{equation}
where $n$ is an integer, and the $k_y$-resolved conductance is given by
\begin{equation}
G'(k_y)=G_0\sum_{\nu'=K,K'}\sum_{\nu=K,K'}T_{\nu',\nu}(k_y).
\label{cond-G-ky}
\end{equation}
Here, $T_{\nu',\nu}(k_y)$ denotes the transmission probability from the $\nu$ valley in the left monolayer to the $\nu'$ valley in the right monolayer for the transverse wave number $k_y$.
We impose periodic boundary conditions in the $y$ direction with width $L_y=3000b$.
Only the $k_y$ modes for which the monolayer leads have propagating states
contribute to the conductance.
The corresponding range is approximately given by
 $|3k_y b| < |2\varepsilon/\gamma_0| \simeq 0.1$.
 Therefore, among the data shown in Tables~III and IV,
  those at $k_y=0$ make the largest contribution to the conductance shown in Fig.~5(a).
 The number of allowed $k_y$ modes is 31 in each valley, resulting in a total of 62 transport channels.
The inset of Fig.~5(a) shows that $G'(k_y)$ varies smoothly over the allowed discrete values of $k_y$
for $E=0$, $\vec{\delta}=(0.1,0)$~\AA, and $N=50$.
For $N<10$, the barrier becomes extremely thin, and the conductance $G$ approaches its maximum value $62G_0$.

Open squares, circles, triangles, and filled squares represent the cases $\vec{\delta}=(0,0)$, $(0,0.1)$, $(0,-0.1)$, and $(0.1,0)$~\AA, respectively.
Because of the period-3 oscillations originating from $k_x^{\mathrm{r}}a\simeq \pm 2\pi/3$, the data are grouped according to $N \bmod 3$. Here, $N \bmod 3$ denotes the remainder when $N$ is divided by 3.
For $N \bmod 3 = 2$ and $N \bmod 3 = 0$, the data are vertically shifted by $20G_0$  and $40G_0$, respectively. 
The displacement-induced change in the conductance, $\Delta G$, is obtained from the difference relative to the open-square data.
 The close agreement between panels~(a) and (b) demonstrates that both $G$ and $\Delta G$ are well described by the approximate Bloch factor  
 $\lambda$ obtained from Eq.~(\ref{DK-eq}).
 The dependence of $\Delta G$ on $\vec{\delta}$ is consistent with 
the behavior of $\Delta k_x^{\mathrm{i}}$ in Table~IV
 for $N>60$. Specifically, for $\vec{\delta}=(0,\pm 0.1)$~\AA, the signs of $\delta_y$, $\Delta |\lambda|$, and $\Delta G$ are consistent, whereas $\Delta G$ is relatively small for $\vec{\delta}=(0.1,0)$~\AA.
However, for $N<50$, the relation between $\Delta G$ and $\delta_y$
tends to be opposite to that observed for larger $N$.
This sign reversal is reproduced in Fig.~5(c).
 When $N$ is small, $|\lambda|^N$ remains close to unity, and the oscillatory effect originating from the phase of $\lambda$ dominates over the attenuation effect.
When $70 < N < 80$ and $\delta_x = 0$, on the other hand,  Figs.~5(b) and~5(c) exhibit the same sign of $\Delta G/\delta_y$, indicating that the effects of the real and imaginary parts of the longitudinal wave number, $k_x^{\mathrm{r}}$ and $k_x^{\mathrm{i}}$, reinforce each other.
As shown in Table~IV, an increase in $\delta_y$ leads to an increase in $|2\xi^{\mathrm{r}}| \simeq  \sqrt{3}|k_x^{\mathrm{r}}a \mp 2\pi/3|$, 
 resulting in a shorter oscillation period of $G$ with respect to $N$.

When $N$ and $\varepsilon$ are sufficiently large,
 Ref.~\cite{Tamura-2025} derived the following asymptotic form
for the conductance:
\begin{equation}
 G'(k_y) \simeq
 F\left|
 \sum_{\tau,l}
 \lambda_{\tau,l}^{2N}\beta_l^{-1}
 \right|^2,
\label{G-asymptotic}
\end{equation}
where
\begin{equation}
\beta_l=
\frac{
2\varepsilon E-ilq\gamma_0^2
}{
\gamma_1(E-\varepsilon)
},
\label{beta}
\end{equation}
and $\lambda_{\tau,l}$ is defined by Eq.~(\ref{lambda-xi})
with $\sigma=+$ and $\xi=\xi_{\tau,l}$.
Here, Eq.~(\ref{beta}) represents the interlayer
wave-function ratio, which will be discussed in
Sec.~VI, and  the prefactor $F$ is independent of $N$.
Except for the phase shift arising from $\beta_l^{-1}$,
the $N$ dependence in Eq.~(\ref{G-asymptotic})
originates from the factors
$\lambda_{\tau,l}^{2N}
=
\exp(iNk_xa)$, as intuitively expected.
The physical role of the armchair edges
was clarified in Ref.~\cite{Tamura-2025}
by comparing MBMGJ and MBMGJ$'$.
Here, MBMGJ denotes the junction discussed in the
present paper, whereas the side view of MBMGJ$'$
is shown in the inset of Fig.~5(b) in a form
similar to that in Fig.~1(a).
In MBMGJ, both armchair edges belong to the
$\downarrow$ layer, whereas in MBMGJ$'$ each layer
contains one armchair edge.
 In both junctions, the ideal armchair edges
and the well-defined value of $N$ give rise to
the factor $\lambda^N$ in
Eq.~(\ref{G-asymptotic}).
In the case of MBMGJ$'$, however, the factor
$\beta_l^{-1}$ in
Eq.~(\ref{G-asymptotic}) disappears.
This behavior was explained using chiral and
rotational operations.

Although the armchair edge is known to be the most stable graphene edge structure, the monolayer--bilayer interfaces realized in experiments may deviate from an ideal armchair configuration.
For example, defects in which carbon dimers partially attach to the interface would locally increase the effective value of $N$.
However, as shown in the figures, the characteristic positive response of the conductance $G$ to $\delta_y$ becomes essentially independent of 
$N \bmod 3$ in the regime where the effects of $k_x^{\mathrm{r}}$ and $k_x^{\mathrm{i}}$ reinforce each other.
Therefore, the influence of such local defects on the conductance is expected to be minor within the parameter range considered here.
In realistic systems, the transverse width $3N_yb$ is finite, which inevitably leads to the formation of edges parallel to the $x$ axis, such as zigzag edges.
These edges can host edge bands inside the gap.
Since the number of such edge bands is of order unity and does not scale with $N_y$, their contribution to the conductance is expected to be on the order of  $G_0$.
In contrast, the contribution of the evanescent states considered here to the conductance scales linearly with $N_y$.
For sufficiently large $N_y$, this contribution therefore dominates over that from the edge bands.
 Indeed, for $N_y=1000$, as used in Fig.~5, the change in conductance 
$\Delta G$ already exceeds $G_0$.

\section{$\widetilde{\Delta}E$ approximation}
In the perturbative calculation presented in this section,
the unperturbed solution consists of the Bloch factor $\lambda$,
 obtained from Eqs.~(\ref{lambda-xi}) and~(\ref{xi-zero}), 
and the corresponding wave function
\begin{equation}
\vec{\psi}_{\tau,l}=
\left(
\begin{array}{c}
\frac{(E+\varepsilon)\omega}{|\gamma_0|\left(\tau\sqrt{p+ilq-s^2 }-i s\right)}
\\
1
\\
\rho_{\tau,l}\beta_l \frac{(E-\varepsilon)\omega^*}{|\gamma_0|\left(\tau\sqrt{p+ilq-s^2 }+i s\right)}
\\
\rho_{\tau,l}\beta_l
\end{array}
\right),
\label{phi}
\end{equation}
 where
\begin{equation}
\rho_{\tau,l}=\omega^2\frac{\tau\sqrt{p-s^2 +ilq}+is}{\tau\sqrt{p-s^2 +ilq}-is},
\end{equation}
and $\beta_l$ is defined by Eq.~(\ref{beta}).
We define the perturbation $H_{\gamma_3}$ by retaining only the $f_3^{(\pm)}$ terms in Eq.~(\ref{def-H}) and setting all other terms to zero.
As discussed in Sec. IV, the effect of $\gamma_4$
can be neglected in the vicinity of $E=0$.
Therefore, we set $\gamma_4 = 0$  in $H_{\gamma_3}$.
The energy shift induced by $H_{\gamma_3}$ is then given,
within first-order perturbation theory, by
\begin{equation}
\widetilde{\Delta}E_{\tau,l}
=
\frac{{}^{t}\vec{\psi}^{\,*}_{\tau,l}
\, H_{\gamma_3}\,
\vec{\psi}_{\tau,l}}
{\left|\vec{\psi}_{\tau,l}\right|^2}.
\label{dE-1}
\end{equation}
We refer to this approximation as the $\widetilde{\Delta}E$ approximation.
The RDL is obtained as follows.
(i) The value $\xi^{(0)}$ at $E=E'$ is calculated from Eq.~(\ref{xi-zero}). Using the relation $\xi^{(0)}-1=2\cos(k_x a/2)$, we obtain
the RDL data $(k_x a, E')$ for the only-$\gamma_1$ model.
(ii) Substituting $E=E'$ into Eq.~(\ref{dE-1}), we obtain the RDL data $(k_x a, E'+\widetilde{\Delta}E_{\tau,l})$ for the $\widetilde{\Delta}E$ approximation.
The same set of $E'$ values is used in both procedures, and 
 $q$ is restricted to be purely imaginary in Eq.~(\ref{def-q}), so that $iq = |q|$.
For the RDL, both $k_x$ and $E$ are real quantities. Consequently, one may either fix $E$ and evaluate the change in $\xi$ within the $\widetilde{\Delta}\xi$ approximation, or fix $k_x$ and evaluate the energy shift within the $\widetilde{\Delta}E$ approximation.
For the CDL, however, $k_x$ generally has both real and imaginary parts, and it is not possible to fix them simultaneously.
As a result, the $\widetilde{\Delta}E$ approximation, which requires $k_x$ to be fixed at a real value, cannot be applied to the CDL.
For the CDL, $q$ is real and $H_{\gamma_3}$ is non-Hermitian;
consequently, Eq.~(\ref{dE-1}) becomes complex.
The $\widetilde{\Delta}\xi$ approximation is applicable to both the RDL and the CDL, but it breaks down near the band edge of the gap, where $q\to 0$.
In contrast, although the $\widetilde{\Delta}E$ approximation is restricted to the RDL, it remains valid even at the gap edge and is therefore complementary to the $\widetilde{\Delta}\xi$ approximation.

\begin{figure}
\begin{center}
\includegraphics[width=\linewidth]{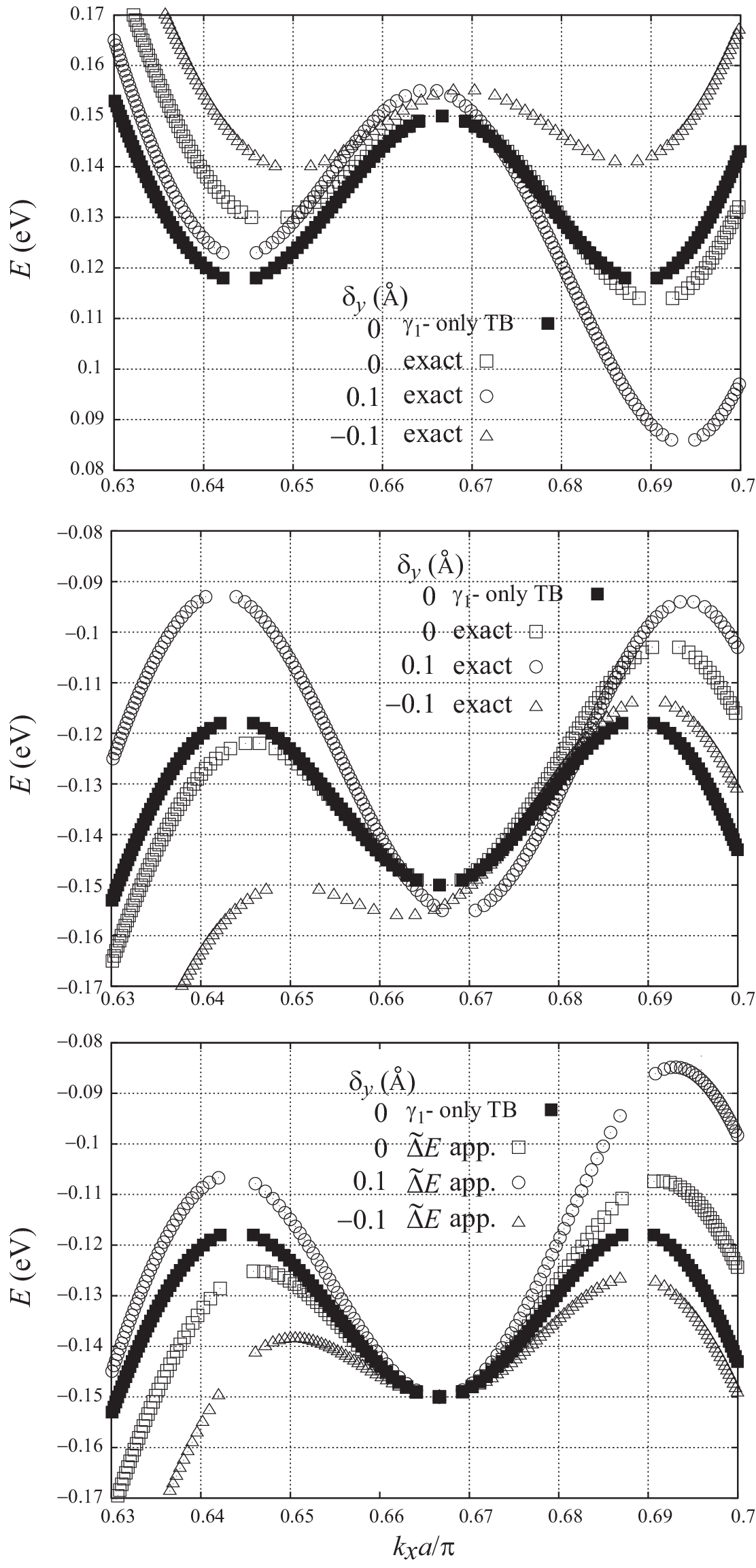}
\caption{
The RDL for $k_y=0$ and $\varepsilon=0.15$~eV.
Except for the filled square symbols, the top, middle, and bottom panels
show the exact $E_3$, the exact $E_2$, and the approximate values
$E' + \widetilde{\Delta}E_{\tau,l}$ corresponding to $E_2$, respectively.
The approximate results are evaluated using an $E'$ spacing of
$0.001~\mathrm{eV}$.
The exact RDL is calculated as a function of $E$ using the same energy spacing
of $0.001~\mathrm{eV}$.
See the text for the definition of $E'$.
}
\end{center}
\label{Fig6}
\end{figure}
\begin{figure}
\begin{center}
\includegraphics[width=\linewidth]{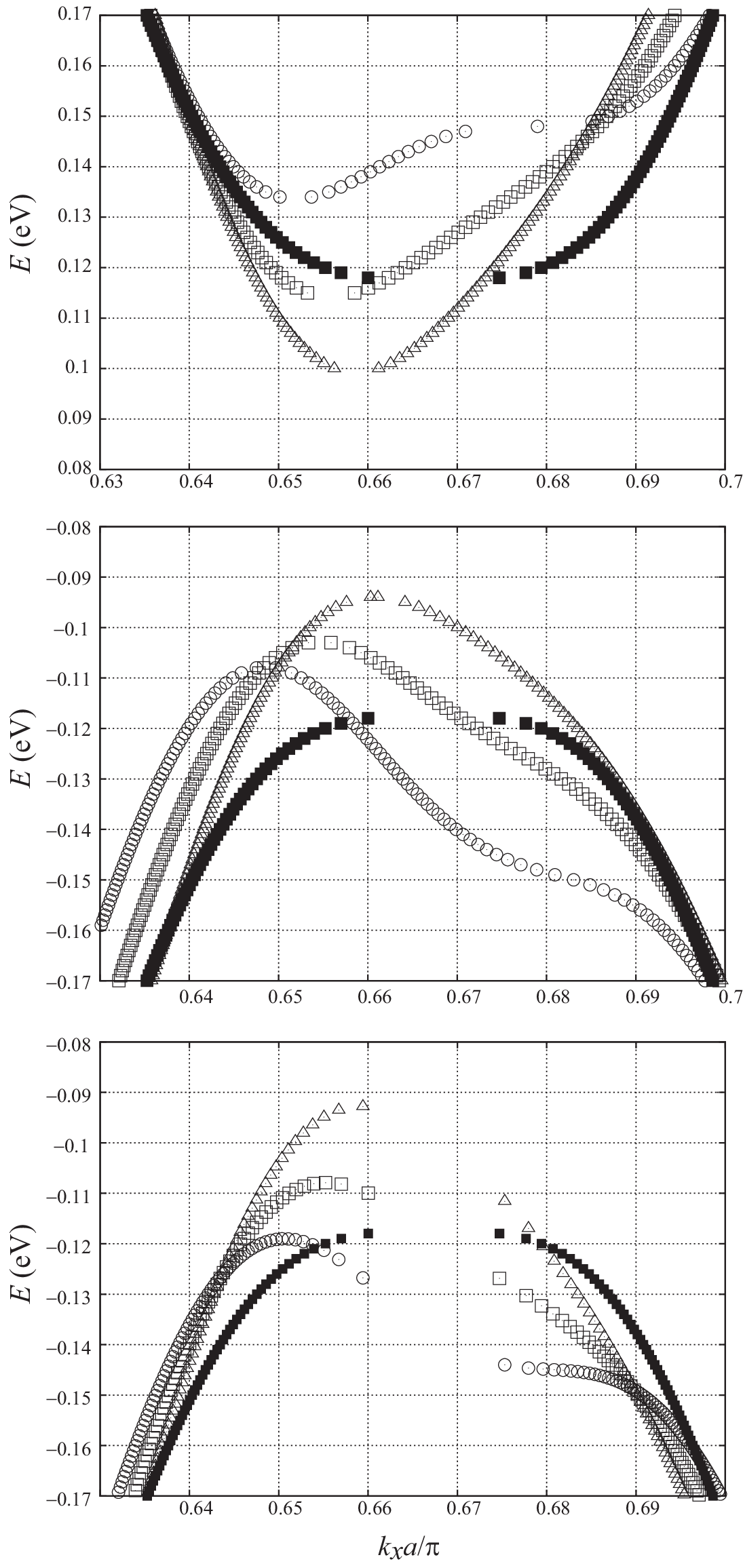}
\caption{
The RDL for $3k_y b = 0.04\pi$ and $\varepsilon=0.15$~eV.
The panel arrangement, the $E'$ spacing used for the approximate results,
the energy spacing used for the exact calculations,
and the legend are the same as in Fig.~6.
}
\end{center}
\label{Fig7}
\end{figure}

As shown in Sec.~V, the effect of $\delta_x$ was found to be small;
we therefore focus on the case $\delta_x = 0$.
Figure~6 shows the RDL for $k_y = 0$, while Fig.~7 shows the RDL for $3k_y b = 0.04\pi$.
Filled square symbols represent the RDL data $(k_x a, E')$.
All other symbols correspond to the RDL in the presence of skew interlayer hopping:
open squares, circles, and triangles represent $\delta_y = 0$, $0.1$~\AA, and $-0.1$~\AA, respectively.
Except for the filled square symbols, the top, middle, and bottom panels depict
the exact $E_3$, the exact $E_2$, and the approximate values
$E' + \widetilde{\Delta}E_{\tau,l}$ corresponding to $E_2$.
The filled and open square symbols correspond to $\gamma_1=\gamma_1^{(0)}$, whereas the circles and triangles correspond to $\gamma_1\neq\gamma_1^{(0)}$.
However, the $k_x a$ data set obtained in procedure (i) hardly depends on
$\delta_y$.
The dependence on $\delta_y$ is extremely small  but still barely visible
around $k_x a = 0.66\pi$ in the bottom panel of Fig.~7.
The exact values of $k_x a$ are obtained from the eigenvalues of the transfer matrix,
as explained in Appendix~B.
For a consistent comparison, the exact calculations are carried out
as a function of $E$ using the same numerical grid
as that adopted for $E'$ in the $\widetilde{\Delta}E$ approximation.
Within the $\widetilde{\Delta}E$ approximation, $E_3(k_x) = -E_2(k_x)$,
and we therefore omit the presentation of the positive-energy branch.
Although the filled square symbols are also symmetric with respect to energy,
they are retained in the top panel in order to illustrate the effect of skew interlayer hopping
in the absence of shear displacement.
In the exact numerical calculations, on the other hand, $\gamma_4$ is retained in the Hamiltonian,
leading to an energy asymmetry.
Since the calculations in procedure (i) are performed as a function of $E'$
with a spacing of $0.001~\mathrm{eV}$, the filled square symbols become sparse
when the RDL is nearly flat.
Consequently, they are absent near
$k_x a = 0.67\pi$ in Fig.~7.
Because the perturbation $H_{\gamma_3}$ does not modify $k_x$ within the
$\widetilde{\Delta}E$ approximation, except for a very small
$\gamma_1-\gamma_1^{(0)}$ effect, the same missing-data behavior is
observed for the open symbols in the bottom panel of Fig.~7.
The difference between the filled and open square symbols is due to the effect of skew interlayer hopping
in the absence of shear displacement, i.e., the $\gamma^{(0)}_{3,4}$ effect.
In contrast, the changes in the triangular and circular symbols relative to the open squares
represent the effect of $\delta_y$, namely the $\gamma_{3,4} - \gamma^{(0)}_{3,4}$ effect.
Although the relative ordering of the four symbols is not necessarily identical
between the exact and approximate results,
the $\gamma^{(0)}_{3,4}$ and $\gamma_{3,4} - \gamma^{(0)}_{3,4}$ effects are in good overall agreement.
The experimentally observable effect is not $\gamma^{(0)}_{3,4}$ itself,
but rather the $\gamma_{3,4} - \gamma^{(0)}_{3,4}$ contribution.
The overall shape of the RDL and the magnitude of the splitting
are well reproduced by the $\widetilde{\Delta}E$ approximation.

Comparing the present results with those presented in Tables III and IV for $3k_yb=0$ and $0.04\pi$, 
we find that, in Fig.~6, $|\lambda|$ naturally approaches unity as the gap width decreases.
In Fig.~7, the magnitude of the $\gamma_{3,4}-\gamma^{(0)}_{3,4}$ effect is comparable to that in Fig.~6
but with the opposite sign, 
 whereas such a sign reversal is not observed in the corresponding data in Table~IV.
However, for most of the $k_y$ range that contributes to the junction conductance $G$, the natural relation at $k_y=0$ is preserved.
For $k_y=0$, $\Delta E_{\tau,l}= \widetilde{\Delta}E_{\tau,l}(\delta_y\neq 0)-\widetilde{\Delta}E_{\tau,l}(\delta_y= 0)$  reduces to a simple form:
 \begin{equation}
\Delta E_{\tau,l} \simeq  \frac{4|\gamma_0^2q|\beta_l\theta}{\gamma_1\left |\vec{\psi}_{\tau,l}^{(0)}\right|^2},
\label{simple-DE}
\end{equation} 
 where $\vec{\psi}_{\tau,l}^{(0)}= \vec{\psi}_{\tau,l}\left. \right|_{k_y=0}$. 
 The interlayer wave-function ratio $\beta_l$ appearing in Eq.~(\ref{simple-DE}) shows, through Eq.~(\ref{dE-1}), how the $\gamma_3$-mediated interlayer coupling contributes to the energy shift.
 Since $iq=|q|$ is close to zero and $|E|<|\varepsilon|$, Eq.~(\ref{beta}) indicates that $\beta_l$ and $E$ have opposite signs. Since $\theta$ increases with increasing $\delta_y$, Eq.~(\ref{simple-DE}) indicates that $\Delta E_{\tau,l}$ has the opposite sign to $E\delta_y$.
  This result is consistent with Ref.~\cite{shear-theory-3} and indicates that variations in the $\gamma_3$-induced interlayer coupling are responsible for the changes in both the band-gap width and $|\lambda|$.

\section{in-plane sublattice wave-function ratios}

In this section, we analyze the quantities $A_\downarrow/B_\downarrow$ and $A_\uparrow/B_\uparrow$, i.e., in-plane sublattice wave-function ratios (ISWRs).
Each ISWR, which is often interpreted as a sublattice pseudospin, provides a compact description of the internal phase structure~\cite{pseudo-spin-1,pseudo-spin-2}.
Although the ISWR  is not directly observable, it offers valuable insight into the structure of decaying modes.
Since the effects of $\delta_x$, $k_y$, and $\gamma_4$ were found to be small,
we first introduce the Hamiltonian matrix $H'$ by setting these parameters to zero.
We then define $H''$ by additionally setting $\gamma_3=0$ in $H'$.
The same prime notation is used for the corresponding wave functions
$(A_\downarrow,B_\downarrow,A_\uparrow,B_\uparrow)$ and for $\xi$;
namely, single and double primes correspond to $H'$ and $H''$,
respectively.
From the secular equation of $H''$, we obtain
\begin{equation}
\left(E-\mathrm{sgn}(\diamond) \varepsilon-\gamma_1\frac{A_{-\diamond}''}{A_{\diamond}''}\right)\frac{A_\diamond''}{B_\diamond''}=-\xi''|\gamma_0|,
\label{H''-secular}
\end{equation}
where $\xi''=\left. \xi^{(0)}\right|_{k_y=0}=-\tau\sqrt{p+i l q}$~\cite{note2}.
 $\diamond$ denotes either $\downarrow$ or $\uparrow$, and 
$\mathrm{sgn}(\downarrow)=-1$, $\mathrm{sgn}(\uparrow)=+1$,
$-\!\!\downarrow=\uparrow$, and $-\!\!\uparrow=\downarrow$.

From Eq.~(\ref{phi}), we obtain
\begin{equation}
\frac{A_\uparrow''}{A_\downarrow''}
=\beta_l\frac{E-\varepsilon}{E+\varepsilon},
\end{equation}
and 
\begin{equation}
\frac{A_\diamond''}{B_\diamond''} =
-\frac{E- \mathrm{sgn}(\diamond)\varepsilon}{|\gamma_0|(p+ilq)}\xi''.
\label{AB-ratio-1}
\end{equation}
By applying the replacements
$A_\diamond''/B_\diamond''\rightarrow A_\diamond'/B_\diamond'$ and
$\xi''\rightarrow\xi'$ in Eq.~(\ref{AB-ratio-1}), 
we obtain an approximate formula
\begin{equation}
\frac{A_\diamond'}{B_\diamond'}=
\tau \frac{E- \mathrm{sgn}(\diamond)\varepsilon}{|\gamma_0|\sqrt{p+ilq}}e^{-il\theta},
\label{AB-ratio-2}
\end{equation}
where $\xi'=\xi''+\Delta\xi \simeq \xi''\exp(-i l \theta)$ has been used.
Equation~(\ref{AB-ratio-2}) has the following properties.
(a) For fixed $(\tau,l)$, since $|\varepsilon|>|E|$, the interlayer phase difference is $\pi$.
(b) For fixed $l$ and a given layer, reversing the sign of $\tau$ causes the $\pi$ phase change.
(c) For fixed $\tau$ and a given layer, reversing the sign of $l$ yields the complex conjugate of the ISWR.
Because $k_y=0$, the system preserves time-reversal symmetry, and property (c)  holds.
Equations~(\ref{lambda-xir-xii}) and (\ref{xi-zero})  show that both operations (b) and (c) correspond to valley exchange, while operation (b) is not a complex conjugation.
Although chiral symmetry is present when $\gamma_4=0$, operation (b) does not correspond to a chiral transformation. 
This distinction can be seen from the transformation of the layer amplitudes.
The amplitudes $(A_\downarrow,A_\uparrow)$ change into
$(-A_\downarrow,-A_\uparrow)$ under operation (b),
whereas they change into
$(A_\downarrow,-A_\uparrow)$ under a chiral transformation
~\cite{chiral-1,chiral-2}.
This difference between the valley-exchange operation and the chiral transformation is closely related to valley current reversal~\cite{up-Tamura,Tamura-2025, updown-Tamura,Tamura-exact, VCR-1}.

In addition to the analytical result of Eq.~(\ref{AB-ratio-2}), we also obtained exact ISWRs from the eigenvectors of the transfer matrix, as shown in Appendix~B. We found that the ISWR phase $\phi$ exhibits a clearer dependence on $\delta_y$ than the ISWR magnitude $|A_\diamond/B_\diamond|$. We therefore focus on the phase $\phi$ and omit the presentation of $|A_\diamond/B_\diamond|$.
The inset of Fig.~8  shows the energy dependence of the phase
$\phi_0 \equiv \arg\!\left[\exp(-i\theta)/\sqrt{p+i q}\right]$,
obtained from Eq.~(\ref{AB-ratio-2}) without $\gamma_4$.
Within this approximation, the ISWR phase $\phi$
takes the values $\pm \phi_0$ and $\pi \pm \phi_0$.
The main panels show the exact $\phi$.
In the exact calculations,
the parameter $\gamma_4$ is retained in the Hamiltonian,
leading to an asymmetry in $E$.
Squares, circles, and  triangles correspond to
$\delta_y=0$, $0.1$~\AA, and $-0.1$~\AA, respectively.
Here, open and filled symbols correspond to  the $\downarrow$ and $\uparrow$ layers, respectively.
The vertical arrows indicate the change in the phase, $\Delta\phi$,
induced by $\delta_y$ near zero energy.
Equation~(\ref{AB-ratio-2}) reproduces the exact $\Delta\phi$
over a relatively wide energy range.
For normal incidence with $k_y=0$, 
the phase $\phi$ is locked to either $0$ or $\pi$ 
within the continuum bands, 
whereas inside the gap it is no longer restricted to these values.

\begin{figure}
\begin{center}
\includegraphics[width=\linewidth]{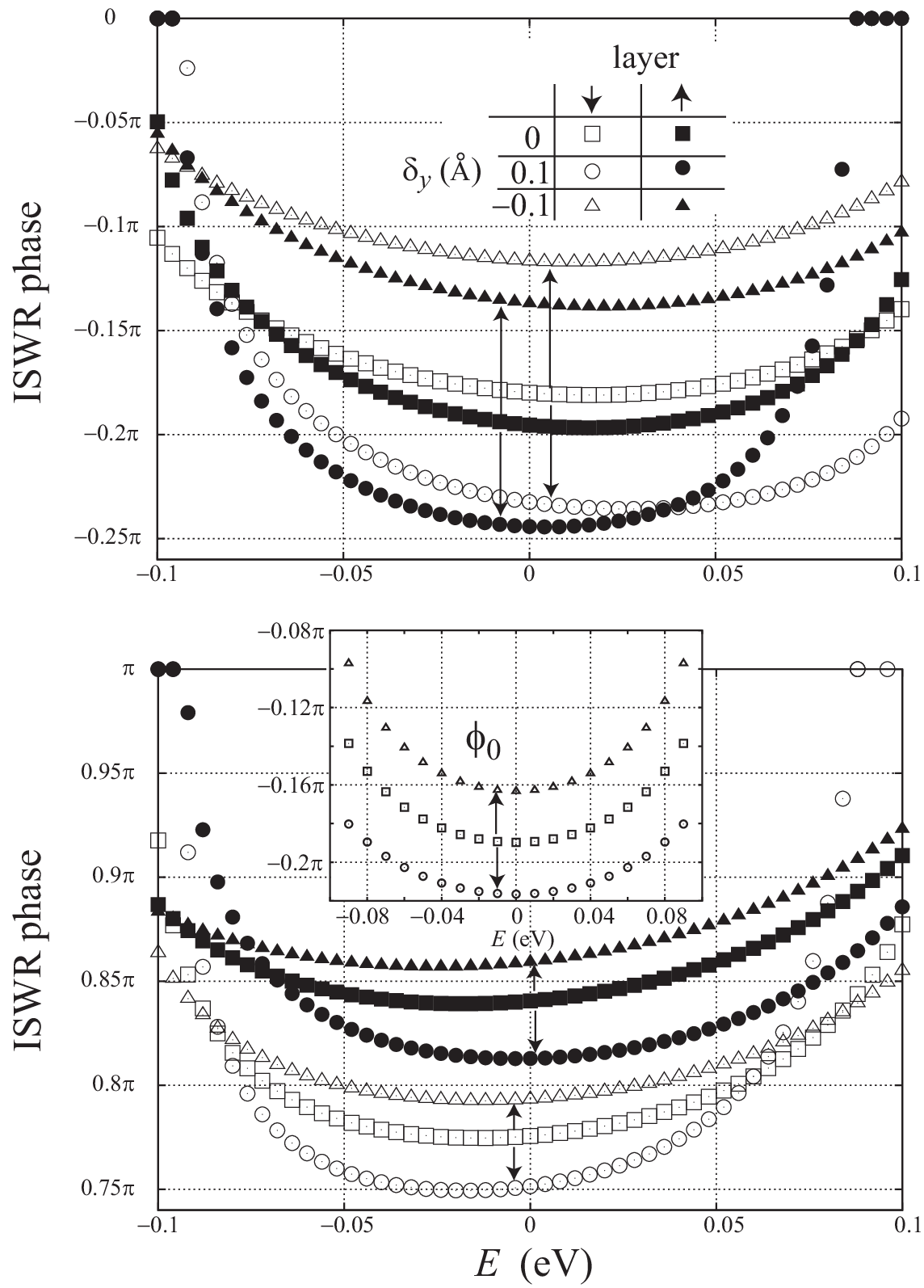}
\caption{
Phase  $\phi$ of the exact ISWR for $\delta_x=0$, $k_y=0$, and
$\varepsilon=0.15$~eV in the ranges
$-0.26\pi < \phi \leq 0$ (top panel) and
$0.74\pi < \phi \leq \pi$ (bottom panel).
Inset: $\phi_0=\arg\!\left[\exp(-i\theta)/\sqrt{p+iq}\right]$.
See the text for the definitions of $\theta$, $p$, and $q$.
}
\end{center}
\label{Fig8}
\end{figure}

\section{Experimental Implications}
The phase and damping time of shear vibrations have been measured using pump--probe optical techniques~\cite{2013NanoLett}.
Such shear vibrations can be driven by pump light with frequencies of several hundred THz.
Moreover, the direction of the shear vibration can be controlled by the polarization direction of the pump light.
 The magnitude of $\vec{\delta}$ used in the present calculations is comparable to the theoretically estimated value of 0.27~\AA\ reported in Ref.~\cite{Rostami}.
Motivated by these experimental findings and the present theoretical results,
we propose an experiment in which the oscillating electric field of
the probe light is converted into a dc current $I_x$, as schematically illustrated in Fig.~1(c).
First, a shear vibration along the $y$ direction is excited by the pump pulse
in the MBMGJ.
On the time scale of the subsequent probe dynamics,
this excitation can be regarded as instantaneous.
Next, a probe pulse with polarization along the $x$ direction, resonant with the shear vibration, arrives with a controlled delay.
The external field associated with the probe light induces an
oscillating probability current $I_x^{(0)}$  in the monolayer regions.
When $I_x^{(0)}$ oscillates in phase with $\delta_y$, electrons move preferentially in the positive $x$ direction during the high-transmission phase corresponding to $\delta_y>0$, resulting in a positive time-averaged current $I_x$.
In contrast, when $I_x^{(0)}$ and $\delta_y$ oscillate with a phase difference of $\pi$, the resulting dc current $I_x$ becomes negative.
The relative phase between $\delta_y$ and $I_x^{(0)}$
can be controlled by the pump--probe delay time.
In the absence of optical excitation, the system is symmetric under
reversal of the $x$ axis. Although the oscillating driving field
instantaneously breaks this symmetry, both the driving field and the
lattice displacement have zero time averages.
The proposed experiment therefore provides a direct test of whether
a dc current can be generated purely by dynamical effects.

The present mechanism is closely related to quantum pumping, in which
a dc current arises from cyclic modulation of system parameters without
an applied bias~\cite{Pump-1,Pump-2,Pump-3,Pump-4}. However, unlike conventional pumping schemes based on
gate voltages or magnetic fields, the dc current proposed here is
generated by controlling the relative phase between the dynamically driven
lattice shear and the probe-induced electronic response.
This distinguishes the present effect from photogalvanic and ratchet
effects, which primarily rely on static spatial asymmetry or optical
nonlinearities~\cite{Ratchet-1,Ratchet-2,Ratchet-3,Ratchet-4,Ratchet-5}.
The present mechanism is also closely related to the phase delay associated
with electron transmission through the bilayer barrier.
Since this phase delay is closely related to various definitions of
tunneling time, the effect may provide an indirect probe of dynamical
aspects of tunneling transport~\cite{tunnel-1,tunnel-2}.

\section{Summary and Perspective }
In summary, we have investigated the evanescent electronic states of Bernal-stacked bilayer graphene under a perpendicular electric field, focusing on low-energy states near the charge neutrality point while extending the analysis to interlayer biases comparable to the
dominant interlayer hopping energy.
Using a full tight-binding model that incorporates the interlayer hopping parameters $\gamma_1$, $\gamma_3$, and $\gamma_4$, we analyzed the Bloch factor $\lambda$ within the electrically induced gap beyond the regime accessible to conventional low-energy analytical approaches.
By approximately solving the secular equation for $\lambda$ in terms of the complex wave number $\xi$, we showed that the $\widetilde{\Delta}\xi$ approximation accurately captures the behavior of evanescent modes over most of the gap, except in the immediate vicinity of the gap edges.
The transport direction was chosen along the zigzag ($x$) direction,
with a generally complex $k_x$ and a real transverse wave number $k_y$.
The solutions were classified according to the
$\lambda$--$(1/\lambda^{*})$ symmetry and the signs of the real and
imaginary parts of $\xi$.

For MBMGJs, we found that the conductance $G$ becomes insensitive to
the displacement $\delta_x$,
 whereas it exhibits a positive response to $\delta_y$, with $dG/d\delta_y>0$ for sufficiently thick bilayer regions.
 Here, the positive direction of $\delta_y$ corresponds to the orientation
in which the angle between the $\gamma_1$ bond and the $\gamma_0$ bond
lying in the $yz$ plane is obtuse.
This behavior is captured by the $\widetilde{\Delta}\xi$ approximation through an analytic connection between the CDL and RDL, while its breakdown near the gap edge is remedied by the complementary $\widetilde{\Delta}E$ approximation.
Together, these results clarify that the effect of $\delta_y$ originates primarily from the $\gamma_3$-induced interlayer coupling.
For the in-plane sublattice wave-function ratio, we also found good agreement between the $\widetilde{\Delta}\xi$ approximation and the exact numerical results.
The phase $\theta$, proportional to $\delta_y$, plays a central role in governing these behaviors.
Based on these findings, we propose an experimental scheme combining a MBMGJ with pump--probe optical techniques to generate a dc current from an optically driven oscillating electric field.
 More broadly, the present results demonstrate that a dc transport response can emerge from the dynamically controlled phase relation between lattice motion and electronic dynamics in evanescent transport regimes,
even in systems that preserve static spatial symmetries.
While the present analysis focuses on bilayer graphene, the analytical
framework developed here can in principle be extended to other layered
two-dimensional materials and to transport along more general
crystallographic directions. Exploring these extensions remains an open
problem for future studies.

\section*{Data availability}

The data and code that support the findings of this article are openly available in the Zenodo repository at \cite{data}.

\appendix

\begin{figure*}
\begin{center}
\includegraphics[width=\linewidth]{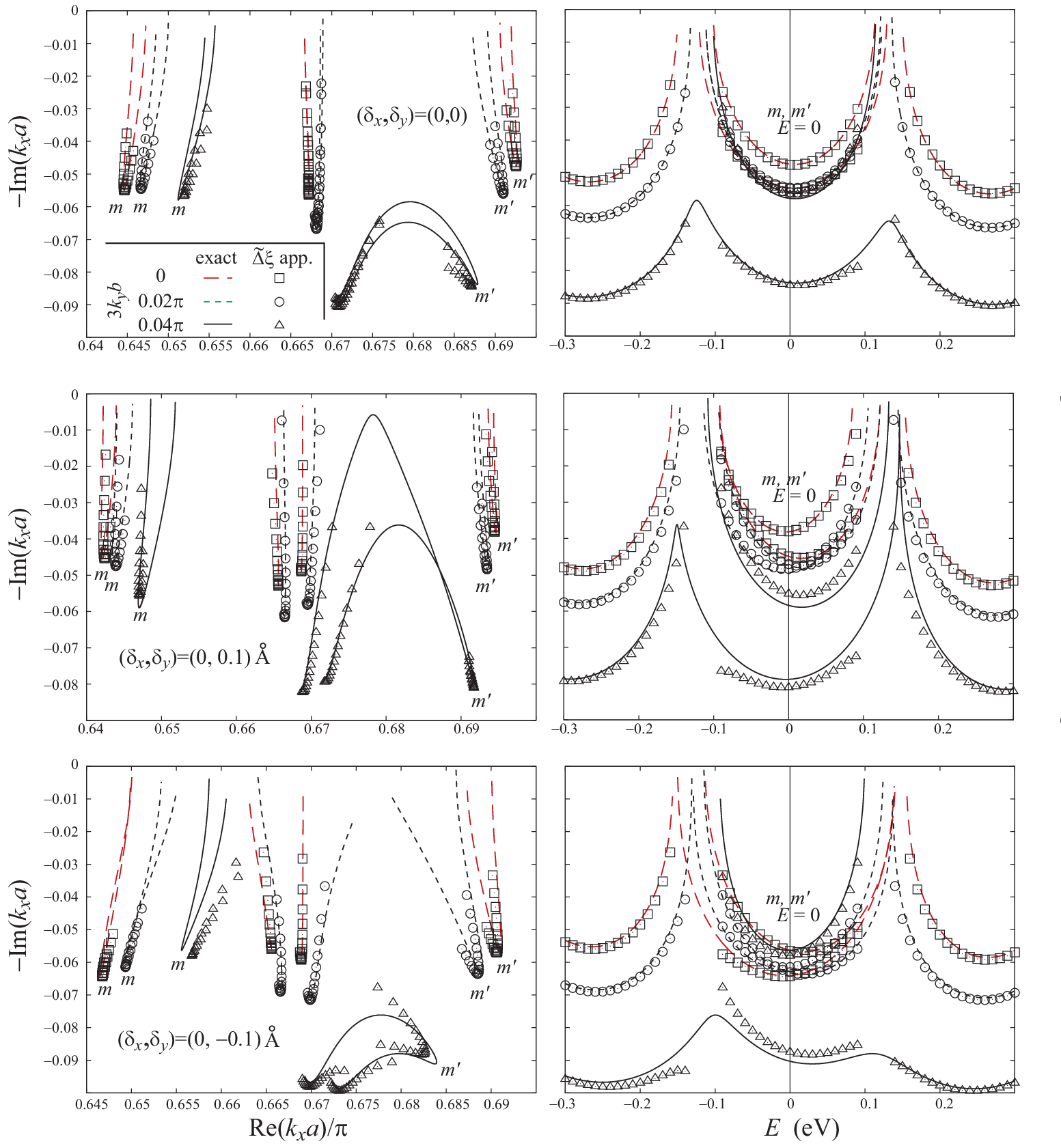}
\caption{
CDLs in the $K$ valley for $\varepsilon=0.15$~eV and three values of $3k_yb$.
The top, middle, and bottom panels correspond to $\vec{\delta}=(0,0)$,
$(0,0.1)$~\AA, and $(0,-0.1)$~\AA, respectively.
Solid, dotted, and dashed lines represent the exact results, whereas
squares, circles, and triangles denote the
$\widetilde{\Delta}\xi$ approximation.
The approximation is evaluated at
$E=-0.3, -0.29, \ldots, 0.3$~eV with an energy spacing of $0.01$~eV,
except in the region where $|q|^2 < 10^{-5}$.
Different line styles and symbols correspond to different values of
$k_y$, as indicated in the legend.
The left and right panels show projections onto the
$(k_x^{\mathrm{r}}a/\pi, -k_x^{\mathrm{i}} a)$ and
$(E, -k_x^{\mathrm{i}} a)$ planes, respectively.
Vertical lines at $E=0$ in the right panels indicate the
correspondence to the points $m$ and $m'$ in the left panels.
}
\end{center}
\end{figure*}
\begin{figure*}
\begin{center}
\includegraphics[width=\linewidth]{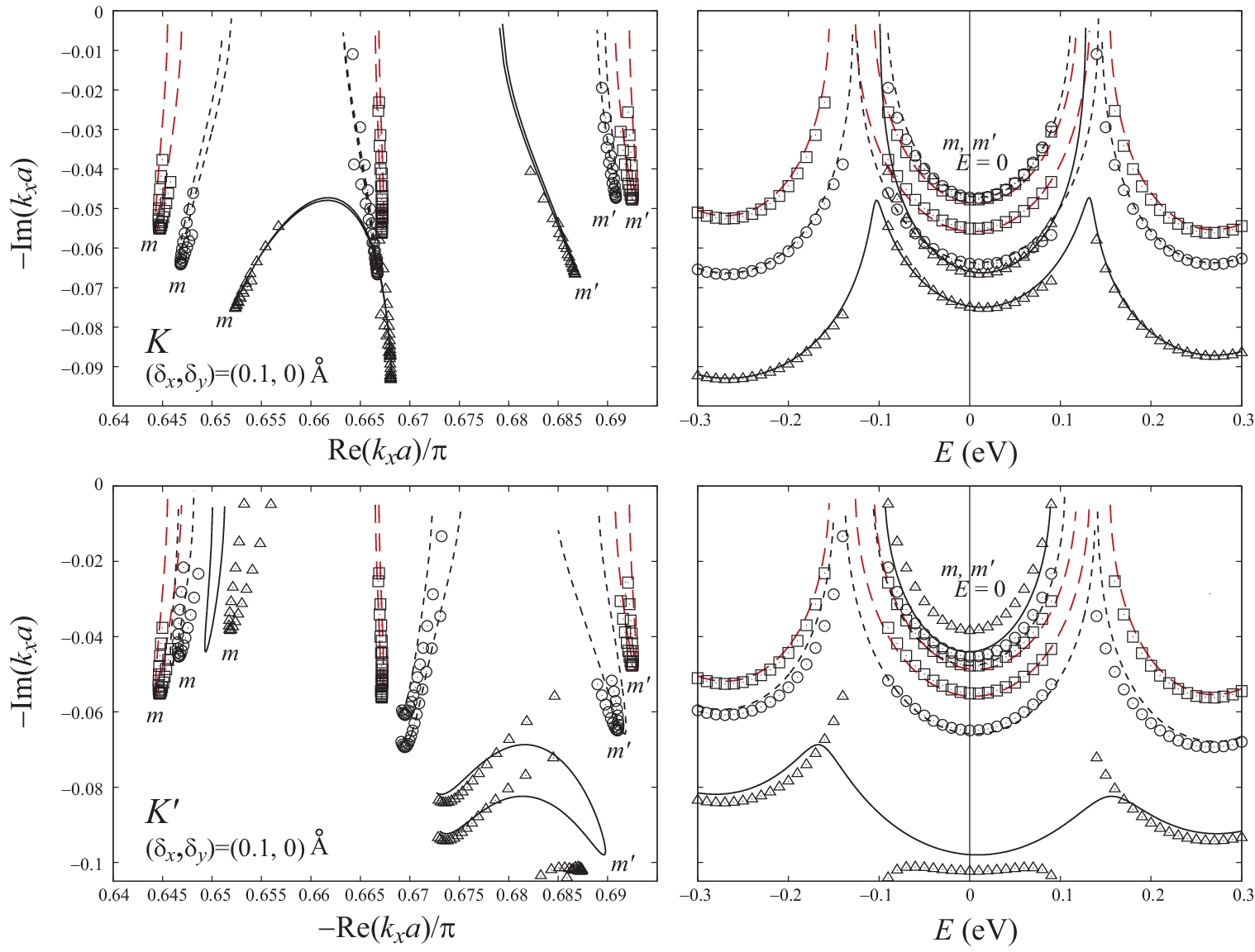}
\caption{
CDLs in the $K$ and $K'$ valleys for
$\varepsilon=0.15$~eV and $\vec{\delta}=(0.1,0)$~\AA.
The top and bottom panels correspond to the $K$ and $K'$ valleys,
respectively.
The projections in the left and right panels and the legend are the
same as in Fig.~9.
In the bottom left panel, the horizontal axis is
$|k_x^{\mathrm{r}}|a/\pi$.
}
\end{center}
\end{figure*}

 \begin{figure}
\includegraphics[width=\linewidth]{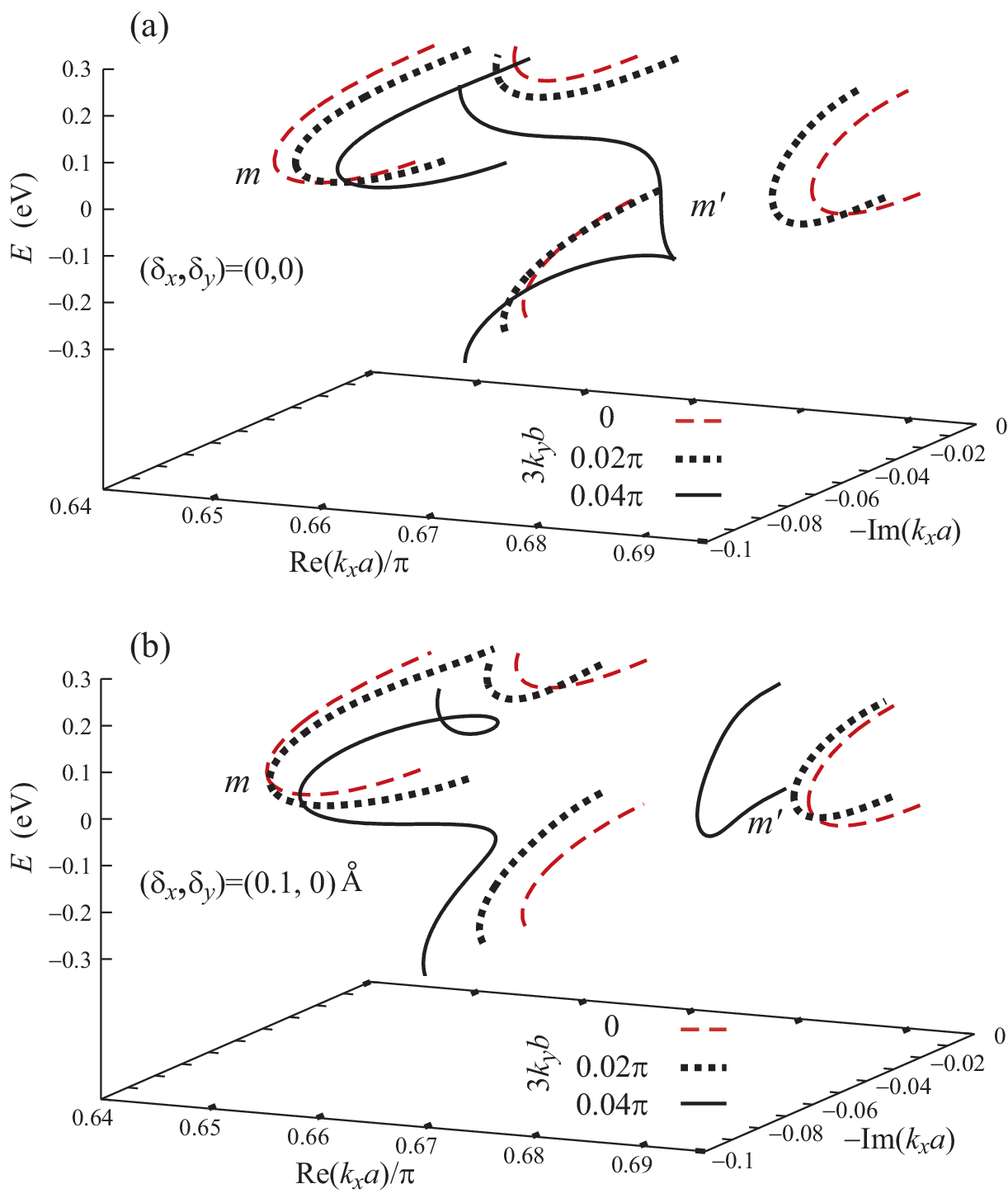}
\caption{
Three-dimensional views of the exact CDLs corresponding to the top
panels of Figs.~9 and~10:
(a) $\vec{\delta}=(0,0)$ and
(b) $\vec{\delta}=(0.1,0)$~\AA.
Dashed, dotted, and solid lines correspond to
$k_y=0$, $3k_yb=0.02\pi$, and $3k_yb=0.04\pi$,
respectively.
}
\label{Fig11}
\end{figure}

 \section{Interdependence of $\xi, \tilde{k}_x,\lambda$, and $k_x$}
In the present paper, we measure the wave number $k_x$ from the $\Gamma$ point. 
In contrast, the effective theory utilizes $\tilde{k}_x$ 
measured relative to the $K$ and $K'$ points ($k_xa=\mp 4\pi/3$).
They are related by
\begin{equation}
k_xa=\tilde{k}_xa\pm  \frac{2}{3}\pi \mp 2\pi,
\end{equation}
where the upper (lower) signs correspond to the $K$ ($K'$) valley.
 When $\tilde{k}_x$ is close to zero, the Bloch factor $\lambda =\exp(ik_xa/2)$ is approximated 
 as
 \begin{eqnarray}
 \lambda &= &e^{\mp i2\pi/3}e^{i\tilde{k}_xa/2}  \label{66-2} \\
   & \simeq & \left(-\frac{1}{2}\mp i \frac{\sqrt{3}}{2}\right)\left(1+i\frac{\tilde{k}_xa}{2}\right)
 \end{eqnarray}
 from which we obtain
\begin{equation}
\xi \simeq \pm \sqrt{3}\tilde{k}_xa/2.
\label{67}
\end{equation}
 
 When $\xi^{\mathrm{i}}=0$,  Eq.~(\ref{lambda-xir}) is trivial.
 We therefore consider the case where  $\xi^{\mathrm{i}} \neq 0$ below.
  Neglecting terms of order $\xi^2$ in  Eq. (\ref{lambda-xi}), we obtain an approximate formula
 \begin{eqnarray}
 \sqrt{\left(\frac{\xi-1}{2}\right)^2-1}  &= & \sqrt{ -\frac{3}{4}-\frac{\xi}{2}+\frac{\xi^2}{4} } \nonumber \\
   & \simeq & \frac{\sqrt{3}}{2} \sqrt{-1-\frac{2}{3}\xi^{\mathrm{r}}-\frac{2}{3}i \xi^{\mathrm{i}}}.
   \label{app-1}
 \end{eqnarray}
  For the square root
\begin{equation}
\sqrt{A}=\exp(i\varphi/2)\sqrt{|A|}
\end{equation}
 of a complex number $A=|A|\exp(i\varphi)$,
we choose the phase $\varphi$ to satisfy
$-\pi<\varphi\le\pi$.
 Under this standard definition, 
\begin{equation}
 \sqrt{-1+o_1 + io_2} \simeq \frac{|o_2|}{2}+\mathrm{sgn}(o_2) i\left(1 -\frac{o_1}{2}\right)
 \label{app-2}
 \end{equation}
 holds for real numbers $o_1$ and $o_2$ satisfying
$|o_1|, |o_2| \ll 1$.
 Using Eqs.~(\ref{app-1}) and  (\ref{app-2}) in Eq.~(\ref{lambda-xi}), we obtain Eq. (\ref{lambda-xir-xii}).
Substitution of Eq. (\ref{67}) into Eq. (\ref{66-2}) yields
\begin{equation}
\lambda=e^{\mp i \left( \frac{2}{3}\pi-\frac{\xi^{\mathrm{r}}}{\sqrt{3}}\right)} e^{\mp \frac{\xi^{\mathrm{i}}}{\sqrt{3}}}.
\label{70}
\end{equation}
Equation~(\ref{lambda-xir-xii}) reduces to Eq.~(\ref{70})
 when $\sigma \mathrm{sgn}(\xi^{\mathrm{i}})$
is chosen to be positive in the $K$ valley
and negative in the $K'$ valley.
\section{Numerical methods}
In this Appendix, we summarize the method of the exact numerical calculation,
which is identical to that of Ref.~\cite{Tamura-exact},
except that the formulation has been rewritten
to suit the present work.
We denote the central bilayer region by ${\rm b}$.
The tight-binding equations
 can be written as
\begin{equation}
0=
h^{(-1)}_\downarrow
\vec{d}_{j-1}^{\;(\downarrow)} +
h^{(0)}_\downarrow
\vec{d}_{j}^{\;(\downarrow)} +
h^{(+1)}_\downarrow
\vec{d}_{j+1}^{\;(\downarrow)},
\label{app-lead}
\end{equation}
for the lead regions ($j \leq -1, N+1 \leq j$),
\begin{equation}
0=
h^{(-1)}_{\rm b}
\vec{d}_{j-1}^{\;({\rm b})} +
h^{(0)}_{\rm b}
\vec{d}_{j}^{\;({\rm b})} +
h^{(+1)}_{\rm b}
\vec{d}_{j+1}^{\;({\rm b})},
\label{app-bi}
\end{equation}
for the bilayer region ($2 \leq j \leq N-2$),
where
$\;^t\vec{d}_{j}^{\;({\rm b})}
=
\left(
^t\vec{d}_{j}^{\;(\downarrow)},
\;^t\vec{d}_{j}^{\;(\uparrow)}
\right)$,
and
\begin{equation}
h^{(\varsigma)}_{\rm b}
=
\left(
\begin{array}{cc}
h^{(\varsigma)}_\downarrow & W^{(\varsigma)} \\
{}^t\!\left( W^{(-\varsigma)} \right)^* & h^{(\varsigma)}_\uparrow
\end{array}
\right).
\label{h-bi}
\end{equation}
Here, $\varsigma=0, \pm 1$, and $W$ is determined by the interlayer hopping.
In addition,
$h^{(-1)}_\downarrow = h^{(+1)}_\downarrow$
and
$h^{(-1)}_\uparrow = h^{(+1)}_\uparrow$.
When $\delta_x \neq 0$, however,
$W^{(-1)} \neq W^{(+1)}$.
The diagonal elements of
$h_\downarrow^{(0)}$
and
$h_\uparrow^{(0)}$
include $-E$, and thus
the relation between Eqs.~(\ref{def-H}) and (\ref{h-bi}) is given by
\begin{equation}
H =
\frac{1}{\lambda} h_{\rm b}^{(-1)}
+
(h_{\rm b}^{(0)}+E{\bf 1}_4)
+
\lambda h_{\rm b}^{(+1)} .
\end{equation}
Since $k_y$ is fixed throughout the calculation,
each $h_{\rm b}^{(\varsigma)}$ is represented by a $4 \times 4$ matrix,
while each $h_{\downarrow}^{(\varsigma)}$ is a $2 \times 2$ matrix.

For the armchair boundaries shown in Fig.~1(a),
the tight-binding equations become
\begin{equation}
0=
h^{(-1)}_\downarrow
\vec{d}_{-1}^{\;(\downarrow)} +
h^{(0)}_\downarrow
\vec{d}_{0}^{\;(\downarrow)} +
\left(
\begin{array}{cc}
h^{(+1)}_\downarrow & W^{(+1)}
\end{array}
\right)
\left(
\begin{array}{c}
\vec{d}_{1}^{\;(\downarrow)}
\\
\vec{d}_{1}^{\;(\uparrow)}
\end{array}
\right),
\label{app-0}
\end{equation}
\begin{equation}
0=
h^{(-1)}_{\rm b}
\left(
\begin{array}{c}
\vec{d}_{0}^{\;(\downarrow)}
\\
0
\end{array}
\right)
+
h^{(0)}_{\rm b}
\left(
\begin{array}{c}
\vec{d}_{1}^{\;(\downarrow)}
\\
\vec{d}_{1}^{\;(\uparrow)}
\end{array}
\right)
+
h^{(+1)}_{\rm b}
\left(
\begin{array}{c}
\vec{d}_2^{\;(\downarrow)}
\\
\vec{d}_{2}^{\;(\uparrow)}
\end{array}
\right),
\label{app-bound-1}
\end{equation}

\begin{equation}
0=
h^{(-1)}_{\rm b}
\left(
\begin{array}{c}
\vec{d}_{N-2}^{\;(\downarrow)}
\\
\vec{d}_{N-2}^{\;(\uparrow)}
\end{array}
\right)
+
h^{(0)}_{\rm b}
\left(
\begin{array}{c}
\vec{d}_{N-1}^{\;(\downarrow)}
\\
\vec{d}_{N-1}^{\;(\uparrow)}
\end{array}
\right)
+
h^{(+1)}_{\rm b}
\left(
\begin{array}{c}
\vec{d}_N^{\;(\downarrow)}
\\
0
\end{array}
\right),
\label{app-N-1}
\end{equation}
and
\begin{equation}
0=
\left(
\begin{array}{cc}
h^{(-1)}_\downarrow & W^{(-1)}
\end{array}
\right)
\left(
\begin{array}{c}
\vec{d}_{N-1}^{\;(\downarrow)}
\\
\vec{d}_{N-1}^{\;(\uparrow)}
\end{array}
\right)
+
h^{(0)}_\downarrow
\vec{d}_{N}^{\;(\downarrow)}
+
h^{(+1)}_\downarrow
\vec{d}_{N+1}^{\;(\downarrow)} .
\label{app-N}
\end{equation}

As in Sec.~V, the exact Bloch factor
$\exp(i k_x a/2)$
is denoted by $\widetilde{\lambda}$,
while $\lambda$ represents the corresponding quantity obtained within the
$\widetilde{\Delta}\xi$ approximation.
For the monolayer region, the exact Bloch factor
$\widetilde{\Omega}$
is employed.
From Eqs.~(\ref{app-lead}) and (\ref{app-bi}),
the transfer matrix is obtained as
\begin{equation}
\Gamma_\diamond
=
-\left(
\begin{array}{cc}
0 & -{\bf 1}_\diamond \\
\frac{1}{h^{(+1)}_\diamond}h^{(-1)}_\diamond &
\frac{1}{h^{(+1)}_\diamond}h^{(0)}_\diamond
\end{array}
\right),
\label{app-transfer}
\end{equation}
where
$\diamond=\downarrow$ or ${\rm b}$,
${\bf 1}_\downarrow={\bf 1}_2$,
and
${\bf 1}_{\rm b}={\bf 1}_4$.
By diagonalizing this matrix,
the eigenvalues and eigenvectors are obtained as
\begin{equation}
\Gamma_{\rm b}
\left(
\begin{array}{c}
\vec{u}_n \\
\widetilde{\lambda}_n \vec{u}_n
\end{array}
\right)
=
\widetilde{\lambda}_n
\left(
\begin{array}{c}
\vec{u}_n \\
\widetilde{\lambda}_n \vec{u}_n
\end{array}
\right)
\label{app-transfer-bi}
\end{equation}
and
\begin{equation}
\Gamma_{\downarrow}
\left(
\begin{array}{c}
\vec{v}_n^{(\pm)} \\
\widetilde{\Omega}_n^{(\pm)} \vec{v}_n^{(\pm)}
\end{array}
\right)
=
\widetilde{\Omega}_n^{(\pm)}
\left(
\begin{array}{c}
\vec{v}_n^{(\pm)} \\
\widetilde{\Omega}_n^{(\pm)} \vec{v}_n^{(\pm)}
\end{array}
\right).
\label{app-transfer-mono}
\end{equation}
When
$|\widetilde{\Omega}_n^{(\pm)}| = 1$,
the eigenvector of
$\Gamma_\downarrow$
is normalized so that
$\vec{v}^{(+)}$
($\vec{v}^{(-)}$)
carries positive (negative) unit probability current, i.e.,
\begin{equation}
\mathrm{Im}\!\left(
\widetilde{\Omega}_n^{(\pm)\,*}
\,{}^{t}\vec{v}_n^{(\pm)\,*}
h_\downarrow^{(1)}
\vec{v}_n^{(\pm)}
\right)
=
\pm 1.
\end{equation}
The exact value of
$k_x a$
is obtained from
$\exp(i k_x a)=\widetilde{\lambda}_n^2$.

We construct the matrices
\begin{equation}
\left(
\begin{array}{c}
U_\downarrow \\
U_\uparrow
\end{array}
\right)
=
\left(
\vec{u}_1,
\vec{u}_2,
\cdots,
\vec{u}_8
\right),
\end{equation}
\begin{equation}
V_\pm
=
\left(
\vec{v}_1^{(\pm)},
\vec{v}_2^{(\pm)}
\right),
\label{V+}
\end{equation}
\begin{equation}
\Lambda
=
{\rm diag}
(
\widetilde{\lambda}_1,
\widetilde{\lambda}_2,
\cdots,
\widetilde{\lambda}_8
),
\end{equation}
and
\begin{equation}
\Omega_\pm
=
{\rm diag}
(
\widetilde{\Omega}_1^{(\pm)},
\widetilde{\Omega}_2^{(\pm)}
).
\end{equation}
We represent the wave function as
\begin{equation}
\vec{d}_j^{\;(\downarrow)}
=
V_+\Omega_+^j\vec{x}_{+}^{\;\rm L}
+
V_-\Omega_-^j\vec{x}_-^{\;\rm L}
\label{left-d}
\end{equation}
for the left lead region $(j \leq 0)$,

\begin{equation}
\vec{d}_j^{\;(\downarrow)}
=
V_+\Omega_+^{j-N}\vec{x}_{+}^{\;\rm R}
+
V_-\Omega_-^{j-N}\vec{x}_-^{\;\rm R}
\label{right-d}
\end{equation}
for the right lead region $(N \leq j)$, and
\begin{equation}
\left(
\begin{array}{c}
\vec{d}_{j}^{\;(\downarrow)}
\\
\vec{d}_{j}^{\;(\uparrow)}
\end{array}
\right)
=
\left(
\begin{array}{c}
U_\downarrow
\\
U_\uparrow
\end{array}
\right)
\Lambda^j \vec{x}^{\;\rm b}
\label{bi-d}
\end{equation}
for the bilayer region $(1 \leq j \leq N-1)$.

Equations~(\ref{app-transfer-bi}) and (\ref{app-transfer-mono}) guarantee
\begin{equation}
\sum_{\varsigma = -1}^{+1}
h_\downarrow^{(\varsigma)}
\left(
\begin{array}{cc}
V_+\Omega_+^\varsigma &
V_-\Omega_-^\varsigma
\end{array}
\right)
=
\left(
\begin{array}{cc}
0 & 0
\end{array}
\right),
\label{lead-hV}
\end{equation}
and
\begin{equation}
\sum_{\varsigma = -1}^{+1}
h_{\rm b}^{(\varsigma)}
\left(
\begin{array}{c}
U_\downarrow \\
U_\uparrow
\end{array}
\right)
\Lambda^\varsigma
=
\left(
\begin{array}{c}
0 \\
0
\end{array}
\right).
\label{bi-hU}
\end{equation}
Applying Eqs.~(\ref{left-d}), (\ref{right-d}),
(\ref{bi-d}), (\ref{lead-hV}),
and (\ref{bi-hU})
to the boundary conditions
(\ref{app-0}), (\ref{app-bound-1}),
(\ref{app-N-1}), and (\ref{app-N}),
we obtain
\begin{equation}
V_+\Omega_+\vec{x}_+^{\; \rm L}
+
V_-\Omega_-\vec{x}_-^{\; \rm L}
=
g_+\vec{x}^{\;\rm b},
\label{app-cond-1}
\end{equation}
\begin{equation}
\left(
\begin{array}{c}
V_+\vec{x}_+^{\; \rm L}
+
V_-\vec{x}_-^{\; \rm L}
\\
0
\end{array}
\right)
=
\left(
\begin{array}{c}
U_\downarrow \\
U_\uparrow
\end{array}
\right)
\vec{x}^{\;\rm b},
\label{app-cond-2}
\end{equation}
\begin{equation}
\left(
\begin{array}{c}
V_+\vec{x}_+^{\; \rm R}
+
V_-\vec{x}_-^{\; \rm R}
\\
0
\end{array}
\right)
=
\left(
\begin{array}{c}
U_\downarrow \\
U_\uparrow
\end{array}
\right)
\Lambda^N\vec{x}^{\;\rm b},
\label{app-cond-3}
\end{equation}
and
\begin{equation}
V_+\Omega_+^{-1}\vec{x}_+^{\; \rm R}
+
V_-\Omega_-^{-1}\vec{x}_-^{\; \rm R}
=
g_- \Lambda^{N-1}\vec{x}^{\;\rm b},
\label{app-cond-4}
\end{equation}
where
\begin{equation}
g_\pm
=
U_\downarrow
-
\frac{1}{h_\downarrow^{(\pm 1)}}
W^{(\pm 1)}
U_\uparrow .
\end{equation}
Combining
Eqs.~(\ref{app-cond-1}),
(\ref{app-cond-2}),
(\ref{app-cond-3}),
and (\ref{app-cond-4}),
we obtain the matrix equation
\begin{equation}
Y
\left(
\begin{array}{c}
\vec{x}_-^{\; \rm L}\\
\vec{x}_+^{\; \rm R}\\
\vec{x}^{\; \rm b}
\end{array}
\right)
=
Z
\left(
\begin{array}{c}
\vec{x}_+^{\; \rm L}\\
\vec{x}_-^{\; \rm R}
\end{array}
\right),
\end{equation}
with the  12 $\times$ 12 matrix
\begin{equation}
Y=
\left(
\begin{array}{ccc}
-V_-\Omega_- & 0 & g_+\Lambda \\
-V_- & 0 & U_\downarrow \\
0 & 0 & U_\uparrow \\
0 & -V_+ & U_\downarrow\Lambda^N \\
0 & 0 & U_\uparrow\Lambda^N \\
0 & -V_+\Omega_+^{-1} &g_-\Lambda^{N-1}  
\end{array}
\right),
\end{equation}
and the 12 $\times$ 4 matrix
\begin{equation}
Z=
\left(
\begin{array}{cc}
V_+\Omega_+ & 0 \\
V_+ & 0 \\
0 & 0 \\
0 & V_- \\
0 & 0  \\
0 & V_-\Omega_-^{-1} \\
\end{array}
\right).
\end{equation}
The unitary scattering matrix $S$, satisfying
\begin{equation}
\left(
\begin{array}{c}
\vec{x}_-^{\; \rm L}\\
\vec{x}_+^{\; \rm R}
\end{array}
\right)
=
S
\left(
\begin{array}{c}
\vec{x}_+^{\; \rm L}\\
\vec{x}_-^{\; \rm R}
\end{array}
\right)
\end{equation}
is obtained from the upper block of the matrix
$Y^{-1}Z$.

Equation~(\ref{cond-G-ky}) is obtained as
\begin{equation}
G'(k_y)
=
G_0
\sum_{n=1}^2
\sum_{n'=1}^2
\left|
S_{n'+2,n}
\right|^2.
\label{app-G}
\end{equation}

Figure~5(a) shows the exact conductance $G$
calculated in this manner.
Figure~5(b) presents the result obtained by replacing
$\widetilde{\lambda}_n$
according to
\begin{equation}
\widetilde{\lambda}_n
\Rightarrow
\left\{
\begin{array}{ccc}
\lambda_{\tau,l}
&\cdots  &
|\widetilde{\lambda}_n| < 1
\\
1/(\lambda_{\tau,-l}^*)
& \cdots &
|\widetilde{\lambda}_n| > 1
\end{array}
\right.
\end{equation}
while Fig.~5(c) shows the result obtained using the replacement
\begin{equation}
\widetilde{\lambda}_n
\Rightarrow
\left\{
\begin{array}{ccc}
\lambda_{\tau,l} / |\lambda_{\tau,l}|
& \cdots &
|\widetilde{\lambda}_n| < 1
\\
|\lambda_{\tau,-l}| / (\lambda_{\tau,-l}^*)
& \cdots &
|\widetilde{\lambda}_n| > 1
\end{array}
\right.
\end{equation}
where $\lambda_{\tau,l}$ is defined within the
$\widetilde{\Delta}\xi$ approximation.
Here, the correspondence between
$n$
and
$(\tau,l)$
is determined by the assumptions
\begin{equation}
\mathrm{sgn}(\xi^{\mathrm{r}})
=
\mathrm{sgn}(\mathrm{Re}(\xi^{(0)}))
=
-\tau
\end{equation}
\begin{equation}
\mathrm{sgn}(\xi^{\mathrm{i}})
=
\mathrm{sgn}(\mathrm{Im}(\xi^{(0)}))
=
-\tau l
\end{equation}
 together with
 the relation
\begin{equation}
\xi
=
1+\widetilde{\lambda}_n+\frac{1}{\widetilde{\lambda}_n}.
\end{equation}
\vspace{.2cm}

 \section{Validation of the $\widetilde{\Delta} \xi$ approximation}
The top, middle, and bottom panels of Fig.~9 show the CDLs for $\vec{\delta}=(0,0)$, $(0,0.1)$~\AA, and $(0,-0.1)$~\AA\ in the $K$ valley.
The top and bottom panels of Fig.~10 show the CDLs for $\vec{\delta} =(0.1,0)$~\AA\ in the $K$ and $K'$ valleys, respectively.
The left (right) panels show the projections onto the ($|k_x^{\mathrm{r}}|a/\pi$, $-k_x^{\mathrm{i}}a$) [($E$, $-k_x^{\mathrm{i}}a$)] plane.
Calculations within the $\widetilde{\Delta}\xi$ approximation are performed at $E=-0.3, -0.29, \ldots, 0.3$~eV with an energy spacing of $0.01$~eV. 
 The squares, circles, and triangles represent the $\widetilde{\Delta}\xi$ approximation data, while the lines represent the exact results.
As in Fig.~4, the approximate results are not shown when
$|q|^2 < 10^{-5}$.
We choose three values of $3k_yb$: $0$, $0.02\pi$, and $0.04\pi$.
As  in Fig.~4(c), the merged CDLs undergo drifting at $3k_yb=0.04\pi$.
Figures~11(a) and 11(b) show three-dimensional views
corresponding to the top panels of Figs.~9 and~10, respectively.
Comparing the middle and bottom panels with the top panels at the $m$ and $m'$ points in Fig.~9, 
one observes clear $\delta_y$ effects at zero energy. 
In Fig.~4(b), the positive-energy extrema $3'$ and $3''$ disappear, 
whereas the negative-energy extrema $2'$ and $2''$ at $E_2$ remain, 
indicating an energy asymmetry induced by $\gamma_4$. 
However, in Figs.~9 and~10, the spectra around the $m$ and $m'$ points 
are nearly symmetric with respect to energy, suggesting that the effects of $\gamma_4$ are weak. 
This is consistent with Eq.~(\ref{DK-eq}), in which the $\gamma_4$ terms 
are always multiplied by $E$, which is close to zero.
Comparing Fig.~10 with the top panel of Fig.~9,
one finds that the effect of $\delta_x$ almost vanishes at $k_y=0$.
Within the $\widetilde{\Delta}\xi$ approximation,
the dominant contribution of $\delta_x$ arises through $D_\eta$,
which is exactly zero at $k_y=0$.
For $3k_yb=0.04\pi$, the formation of drifting CDLs
becomes valley asymmetric.
In the $K'$ valley, the $m'$DL undergoes merging and drifting,
  as in the case of $\delta_x=0$.
In contrast, in the $K$ valley, the $m$DL exhibits drifting behavior.

\FloatBarrier


\begin{thebibliography}{99}
\bibitem{note2}
The secular equation of $H''$ also yields
$A_{\downarrow}''\xi''|\gamma_0|+(E+\varepsilon)B_{\downarrow}''=0$
and
$A_{\uparrow}''\xi''|\gamma_0|+(E-\varepsilon)B_{\uparrow}''=0$.
Since these equations do not explicitly involve the dominant
interlayer hopping parameter $\gamma_1$, they are less suitable for
discussing the effect of interlayer coupling than
Eq.~(\ref{H''-secular}).
\end{thebibliography}
\end{document}